%% file: main.tex
\documentclass[10pt,conference]{IEEEtran}

\input{macro}

\input{authors}

\begin{document}

\title{Refine After Generation: Toward Correct and Concise Patches in LLM-based Program Repair}

\maketitle

\input{0.abstract}

\input{1.introduction}

\input{2.related_work}

\input{3.dataset_construction}

\input{4.approach}

\input{5.evaluation}

\input{6.discussion}

\input{7.threats_to_validity}

\input{8.conclusion}

\input{acknowledgement}

\bibliographystyle{IEEEtran}
\bibliography{refs}

\end{document}

%% file: macro.tex
\IEEEoverridecommandlockouts
\usepackage{cite}
\usepackage{amsmath,amssymb,amsfonts}
\usepackage{algorithmic}
\usepackage{graphicx}
\usepackage{textcomp}
\usepackage{xcolor}
\usepackage{array}
\usepackage{caption}
\usepackage{multirow}
\usepackage{ragged2e}
\usepackage{colortbl}
\usepackage{makecell}
\usepackage{framed}
\usepackage{array}
\usepackage[normalem]{ulem}
\usepackage{url}

\newcolumntype{C}[1]{>{\centering\arraybackslash}p{#1}}
\def\BibTeX{{\rm B\kern-.05em{\sc i\kern-.025em b}\kern-.08em
    T\kern-.1667em\lower.7ex\hbox{E}\kern-.125emX}}

\definecolor{findgray}{gray}{0.95}
\newenvironment{findingbox}{%
  \def\FrameCommand##1{%
    \setlength{\fboxsep}{6pt}
    \setlength{\fboxrule}{1.5pt}
    \textcolor{black}{\vrule width \fboxrule}
    \colorbox{findgray}{##1}
    \textcolor{black}{\vrule width \fboxrule}
  }%
  \MakeFramed {\advance\hsize-\width \FrameRestore}}%
 {\endMakeFramed}
\newcommand{\findx}[1]{%
  \begin{findingbox}%
    \noindent\em #1%
  \end{findingbox}%
}

\definecolor{codegreen}{rgb}{0,0.6,0}
\definecolor{codered}{rgb}{1,0,0}
\definecolor{codegray}{rgb}{0.5,0.5,0.5}
\definecolor{codepurple}{rgb}{0.58,0,0.82}
\definecolor{backcolour}{rgb}{0.95,0.95,0.92}
\definecolor{lightgray}{gray}{0.9}

\definecolor{DarkOrange}{rgb}{0.8,0.3,0.0}
\definecolor{DarkCyel}{rgb}{1.0, 0.49, 0.0}
\definecolor{yellow-green}{rgb}{0.6, 0.8, 0.2}
\newboolean{showcomments}
\setboolean{showcomments}{true}
\ifthenelse{\boolean{showcomments}}
 { \newcommand{\mynote}[2]{
      \fbox{\bfseries\sffamily\scriptsize#1}
        {\small$\blacktriangleright$\textsf{\emph{#2}}$\blacktriangleleft$}}}
 { \newcommand{\mynote}[2]{}}


%% file: authors.tex
\author{
\IEEEauthorblockN{
    Wenqiang Luo\IEEEauthorrefmark{1},
    Jacky Keung\IEEEauthorrefmark{1},
    Xiaoyu Shi\IEEEauthorrefmark{2},
    Yicheng Sun\IEEEauthorrefmark{1},
    Boyang Yang\IEEEauthorrefmark{3},
    Zhou Yang\IEEEauthorrefmark{4}, and
    Haoye Tian\IEEEauthorrefmark{2}
}
\IEEEauthorblockA{\IEEEauthorrefmark{1}Department of Computer Science, City University of Hong Kong, Hong Kong, China\\
wenqialuo4-c@my.cityu.edu.hk, Jacky.Keung@cityu.edu.hk, yicsun2-c@my.cityu.edu.hk}
\IEEEauthorblockA{\IEEEauthorrefmark{2}Department of Computer Science, Aalto University, Espoo, Finland\\
xiaoyu.shi@aalto.fi, haoye.tian@aalto.fi}
\IEEEauthorblockA{\IEEEauthorrefmark{3}Jisuan Institute of Technology, Beijing, China, buaabarty@gmail.com}
\IEEEauthorblockA{\IEEEauthorrefmark{4}Department of Computing Science, University of Alberta, Edmonton, Canada, zy25@ualberta.ca}
}


%% file: 0.abstract.tex
\begin{abstract}
Large language models (LLMs) have advanced automatic program repair (APR) to the point where agentic systems routinely resolve real-world, repository-level issues. Yet the generated patch, the artifact that validators execute, ranking systems compare, and developers ultimately inspect, has received little scrutiny beyond whether it passes tests.

In this paper, we identify patch verbosity as a major yet overlooked concern in LLM-based APR. Characterizing 28 state-of-the-art approaches on SWE-bench Verified, we find that even successful patches are consistently larger and more complex than developer patches, with the median approach producing 121.78\% more total changes, 80.91\% more net changes, and 43.99\% higher cyclomatic complexity. We further show that this verbosity is rooted in capability-oriented design choices such as iterative refinement and broad context, and can hardly be reduced by surface-level controls such as output format or minimality prompts. Motivated by these findings, we formulate post-generation patch refinement and propose \textbf{RECAP}, a lightweight, plug-and-play adapter that attaches to existing repair frameworks after generation. RECAP's refiner is trained via supervised fine-tuning and direct preference optimization with distilled reasoning traces, on a dataset of patch pairs we construct from multiple sources. Across four host systems, prompting, commit-untangling, and minimality-aware baselines reduce patch size only by sacrificing 49 to 217 resolved instances. In contrast, RECAP achieves a substantially better size-correctness tradeoff, cutting average total changes from +242.14\% to +4.24\% and net changes from +348.24\% to -39.75\% relative to developer patches while preserving or improving resolution by up to 42 instances. Our results indicate that minimality cannot be simply reduced to syntactic compression, and that decoupling minimization from generation offers a practical path to more reviewable repairs.
\end{abstract}

%% file: 1.introduction.tex
\section{Introduction}
\label{sec:introduction}

Large language models (LLMs) have transformed automatic program repair (APR), shifting the field from rigid templates to generative strategies based on prompting, fine-tuning, retrieval-augmentation, and agent-based pipelines \cite{pereira2025systematic, yang2025patch, yang2025survey}.
This shift is critical for repository-level bug fixing, where local file context is insufficient but naive whole-repository inclusion exceeds model limits \cite{chen2024large}.
Modern coding agents address this with multi-step workflows of planning, tool use, and execution feedback \cite{jiang2025agentic}, evaluated by benchmarks like SWE-bench \cite{jimenez2024swe}.
As these pipelines grow mature, the generated patch becomes the central artifact that downstream validators execute, ranking systems compare, and developers inspect.

However, a patch that resolves an issue is not necessarily one developers can trust.
LLM-based APR may produce patches that are textually large, broad in scope, unnecessarily complex, or mixed with indirect and redundant changes.
Such verbosity matters because developers prefer smaller repairs \cite{le2019automated} that are more likely to be accepted \cite{weissgerber2008small}, and industrial evidence shows human fixes are usually compact \cite{rondon2025evaluating}.
Reviewers reject large patches not for size itself but for the higher risk of unnecessary changes \cite{tao2014writing} that raise review confusion, effort, and rejection \cite{ebert2021exploratory}.
Verbose repairs are further linked to fault-proneness \cite{purushothaman2005toward}, regressions and poor localization \cite{mechtaev2015directfix}, and wasted token budget and latency \cite{li2025aligning}.
This growing risk from large patches motivates patch minimization.

Yet this drive for minimality must be balanced against correctness.
Passing tests makes a patch plausible, but weak oracles and under-constrained suites can still admit overfitting \cite{tian2020evaluating}, so patch minimization cannot be treated as purely syntactic compression.
A useful refinement method must reduce unnecessary changes without degrading the correctness needed to resolve the issue. Existing APR research largely targets producing, validating, and selecting candidate patches \cite{yang2025patch,yang2025survey,pereira2025systematic,jiang2025agentic}, paying less attention to a complementary post-generation question: \textbf{once an APR system produces a plausible patch, is that patch already good enough for developers to inspect, maintain, and integrate, or can it be further refined while preserving its correctness?}

Prior works motivate this question but do not answer it.
Commit untangling \cite{li2022utango,zhu2026atomizer} isolates development intents, yet the extracted commit may remain internally complex.
Delta debugging \cite{zhang2025toward,zhou2025wdd} minimizes changes under preserved properties but demands repeated validation runs that become prohibitively expensive for large-scale APR.
Minimality-aware methods like AdaPatcher \cite{dai2025less} and QiMeng-PRepair \cite{ke2026qimeng} embed minimality into generation by training small models, which trail the frontier LLMs that dominate the SWE-bench leaderboard~\cite{swebenchlead} at repository-level repair and remain restricted to function-level code.
Meanwhile, state-of-the-art frameworks rely on massive LLMs and complex multi-agent workflows whose retraining imposes severe computational burden.

\textbf{This work.}
We target post-generation refinement of LLM-generated candidate patches for repository-level issue fixes, integrating with existing systems after their patch-generation stage.
Specifically, we propose \textbf{RECAP} (\textbf{RE}fine for \textbf{C}orrect \textbf{A}nd concise \textbf{P}atches), a lightweight post-generation adapter for plug-and-play patch refinement.
Refinement starts from an existing patch rather than synthesizing one from scratch, which lowers the capability barrier enough for a compact specialized model to be viable compared to direct repair.
Given relevant context and a candidate patch, RECAP attempts to return a more concise refined patch through three components, where the collector curates the context and candidate patch provided by the host framework, the filter decides whether a candidate should be refined under the active deployment policy, and the refiner generates the refined patch.

Training the refiner requires data mapping verbose patches to concise targets.
We construct a 5,540-instance patch-refinement dataset from three complementary sources including function-level repair pairs, tangled commits paired with their atomic fixes, and synthetic SWE-bench instances created by inflating concise developer patches.
We train with supervised fine-tuning (SFT) and direct preference optimization (DPO) \cite{rafailov2023direct}, strengthened by reasoning traces that expose the verbose-to-concise transformation \cite{rafailov2023direct,feng2026empirical,wei2022chain,zhuang2025unicott,yu2025cot}.

We evaluate on SWE-bench Verified along three research questions.
First, we characterize patch verbosity across 28 leaderboard approaches, such as SWE-agent \cite{sweagent}, OpenHands \cite{wang2025openhands}, Agentless \cite{xia2024agentless}, and analyze design factors associated with larger patches.
Second, we apply RECAP to patches from different frameworks and compare against prompting, commit-untangling, and minimality-aware baselines.
Third, we ablate the training phases to evaluate their contribution.

Our results show that patch verbosity is widespread, where across 28 approaches the median successful patch has +121.78\% more total changes, +80.91\% more net changes, and +43.99\% higher cyclomatic complexity than the developer patch, tied to design choices such as iterative refinement and broad context rather than surface-level controls like output format or minimality prompts.
RECAP achieves a substantially better size-correctness tradeoff than baselines, which lose 49 to 217 resolved instances while shrinking patches.
Across four host systems, RECAP cuts average total changes from +242.14\% to +4.24\% and net changes from +348.24\% to -39.75\% relative to developer patches, while preserving or improving resolution by up to +42 instances.
Our ablation shows that SFT supplies the core refinement ability while DPO is essential for preserving correctness lost by aggressive SFT-only simplification.

This paper makes the following contributions:
\begin{itemize}
  \item We empirically characterize patch verbosity in state-of-the-art SWE-bench APR systems, showing that many resolved patches remain larger, broader, or more complex than corresponding developer patches.
  \item We formulate post-generation patch refinement for APR and propose an adapter that can be integrated after the patch-generation stage of existing APR frameworks.
  \item We construct a multi-source patch-refinement dataset with verbose-to-concise pairs, reasoning traces, and preference pairs for training a specialized refiner.
  \item We perform a comprehensive evaluation of RECAP on SWE-bench Verified to reveal the effectiveness of post-generation patch refinement.
\end{itemize}

%% file: 2.related_work.tex
\section{Related Work}
\label{sec:related-work}

\textbf{LLM-based Program Repair.} Recent APR is increasingly driven by LLM-based systems evaluated on repository-level benchmarks such as \textsc{SWE-bench}~\cite{jimenez2024swe}. Some improve the \emph{repair process} through fault localization, context retrieval, and patch generation, search- and retrieval-oriented approaches (e.g., \textsc{AutoCodeRover}~\cite{zhang2024autocoderover}, \textsc{SWE-Search}~\cite{moatless}, \textsc{LingmaAgent}~\cite{lingma}) improve repository exploration, while pipeline-based methods (e.g., \textsc{SWE-Fixer}~\cite{swefixer}) decompose repair into stages such as file retrieval and code editing. A second direction enhances agent capabilities through reinforcement learning or experience reuse (e.g., \textsc{DeepSWE}~\cite{deepswe2025}, \textsc{SWE-Exp}~\cite{sweexp}) and transferable procedural knowledge (e.g., \textsc{Lingxi}~\cite{yang2025lingxi}), whereas \textsc{Agentless}~\cite{xia2024agentless} shows that simplified non-agentic workflows remain competitive. Other work studies scaling (e.g., \textsc{Skywork-SWE}~\cite{zeng2025skywork}), efficiency (e.g., \textsc{PatchPilot}~\cite{li2025patchpilot}), and infrastructure (e.g., \textsc{OpenHands}~\cite{wang2025openhands}). Most systems emphasize process and system efficiency, with limited attention to enforcing simplicity in patch generation.

\textbf{Patch Simplicity and Its Importance.} Empirical studies consistently show that smaller, simpler patches are preferable. Larger patches are more error-prone~\cite{purushothaman2005toward} and less likely to be accepted during code review~\cite{weissgerber2008small}, and developers favor concise changes that avoid unnecessary modifications while preserving correct behavior~\cite{le2019automated}. Complex patches also increase cognitive overhead, as long or tangled changes frequently lead to confusion, lower review quality, and delayed integration~\cite{ebert2021exploratory}. Rejection is often driven not by size alone but by irrelevant or mixed changes within a patch~\cite{tao2014writing}, and in automated repair, overly complex patches further raise the risk of regressions and unintended side effects~\cite{mechtaev2015directfix}. Most real-world fixes are also relatively small, reinforcing the importance of minimal edits~\cite{rondon2025evaluating}.
Despite this, LLM-based APR systems often generate unnecessarily complex patches due to imperfect localization and a tendency to produce self-contained solutions rather than minimal edits~\cite{li2025aligning}. Patch simplicity therefore remains critical for correctness, reviewability, and maintainability, since large or complex patches can degrade both software quality and developer productivity.

\textbf{Towards Simpler and Minimal Patches.} Prior work simplifies code changes in several ways. Commit untangling techniques (e.g., \textsc{Flexeme}~\cite{partachi2020flexeme}, \textsc{UTANGO}~\cite{li2022utango}, \textsc{Atomizer}~\cite{zhu2026atomizer}) decompose mixed commits into coherent units, while delta debugging and program reduction methods (e.g., \textsc{Perses}~\cite{perses}, \textsc{C2D2}~\cite{SongWLc2d2}) isolate failure-inducing changes by iteratively removing irrelevant code while preserving behavior. More recently, minimality has been built into APR, where \textsc{QiMeng-PRepair}~\cite{ke2026qimeng} uses edit-aware rewards to encourage concise patches and AdaPatcher~\cite{dai2025less} applies preference learning to bias repair toward smaller edits. However, these approaches remain loosely integrated with modern LLM-based repair systems, leaving the gap between powerful repair generation and controlled patch simplicity an open challenge.

%% file: 3.dataset_construction.tex
\section{Data Construction}
\label{sec:data-construction}

\subsection{Function-Level Code Pairs}
\label{sec:function-level-data}

Our first source is the ACPR preference dataset \cite{dai2025less}, built around the code consistency rate (CCR), the proportion of code lines preserved after modification.
Each of its 1,227 samples provides two correct solutions whose CCRs differ by at least 0.1, one preserving more original code and one performing a larger rewrite.
We take the lower-CCR solution as the non-minimal repair and the higher-CCR solution as the refinement target.
The input is the problem description, buggy code, and lower-CCR patch, and the target is the higher-CCR patch, teaching refinement at the function level.

\subsection{Tangled CCS Commits}
\label{sec:tangled-commit-data}

Our second source captures repository-level patches where a bug fix is committed together with unrelated maintenance changes.
We use the tangled commit dataset of Koh et al. \cite{koh2026detecting}, built from the manually annotated Conventional Commits Specification (CCS) dataset of Zeng et al. \cite{zeng2025first}, which contains 350 atomic commits each labeled with a single intent such as \texttt{fix}, \texttt{refactor}, or \texttt{build}, mixed into 1,400 multi-label tangled commits.
Selecting tangled commits with at least two CCS labels including \texttt{fix} yields 533 instances, treating the tangled commit as the non-minimal patch and the corresponding atomic \texttt{fix} commit as the target.
Since the dataset provides commit messages rather than issue reports, the atomic fix message serves as the relevant context.

\subsection{Synthetic SWE-bench Data}
\label{sec:synthetic-data}

To improve real-world generalizability, we synthesize data from the SWE-bench training split \cite{jimenez2024swe} with no overlap with the test split, following prior repository-level repair training such as SWE-Llama \cite{jimenez2024swe} and Nemotron-CORTEXA \cite{cortexa}. We inflate concise gold (developer) patches into larger input patches.

\textbf{Candidate pre-processing.}
From roughly 19K training instances, we remove empty-patch, malformed, and over-length cases exceeding a 32K-token budget measured after filling the issue description, repository context, and patch into our prompt template, leaving 11,821 valid instances.

\textbf{Concise-target selection.}
Since not every gold patch is already concise, we use LLM-as-a-Judge \cite{li2024llms} to identify patches that are minimal and free of the inflation patterns in Table~\ref{tab:inflation-rules}, yielding 4,109 concise-target instances.

\textbf{Sampling and training-regime split.}
To control cost, we stratified-sample 50\% of the judged-concise instances while preserving the repository and patch-size distribution (total code changes, as defined in Section~\ref{sec:approach}), giving 2,054 instances, split into SFT and DPO sets at a 4:1 ratio (1,643 SFT, 411 DPO).
The disjoint split reduces SFT memorization during preference tuning, and the larger SFT share follows evidence that SFT remains a primary driver before preference optimization \cite{feng2026empirical}.

\textbf{As-is examples.}
To prevent the model from blindly refining already-concise patches, 20\% of the SFT set are ``as-is'' examples whose input and target are identical gold patches, yielding 328 as-is and 1,315 standard SFT instances.

\begin{table}
\centering
\caption{Patch-inflation rules.}
\label{tab:inflation-rules}
\renewcommand{\arraystretch}{0.9}
\scriptsize
\begin{tabular}{m{0.15\linewidth}|m{0.69\linewidth}}
\hline
\makecell[c]{Category} & \makecell[c]{Inflation patterns} \\
\hline
Overly complex solutions &
Replace simple conditionals with elaborate logic;\newline
Use convoluted, verbose patterns when simpler ones exist;\newline
Over-refactor (e.g., adopt a new framework) when a small fix suffices. \\
\hline
Unnecessary or irrelevant changes &
Touch unrelated code or extra files;\newline
Add needless helpers or wrappers;\newline
Perform multi-site refactors where a local change at the call site would suffice. \\
\hline
Code overwrite &
Rewrite whole functions or large sections;\newline
Duplicate behavior already handled upstream or data already available. \\
\hline
Overfitting fixes &
Handle edge cases explicitly instead of generally;\newline
Add case-by-case checks;\newline
Fix downstream in callers (e.g., \texttt{try/except}) rather than the root cause upstream. \\
\hline
Documen-\newline tation \& comments &
Add docstrings or comments explaining the fix;\newline
Add comments explaining unmodified surrounding code. \\
\hline
Reinventing / bypassing abstractions &
Reimplement built-in parameters, helpers, or framework mechanisms;\newline
Manually transform data covered by existing utilities;\newline
Bypass base classes or default utilities with hardcoded logic. \\
\hline
Formatting / cosmetic noise &
Add whitespace noise (blank lines, indentation);\newline
Make cosmetic-only edits such as renaming or reordering imports;\newline
Add redundant \texttt{else}/\texttt{pass} blocks. \\
\hline
\end{tabular}
\end{table}

\textbf{Constructing verbose input patches.}
For each standard SFT instance, we create large input patches via inflation and rejection sampling.
(1) In inflation, a high-performance closed-source LLM rewrites the gold patch into a larger one while preserving the fix, guided by prior observations of how patches become unnecessarily large or indirect, including overly complex fixes \cite{li2025hybrid,liu2025empirical}, unnecessary or irrelevant edits \cite{alomar2025chatgpt}, code overwriting \cite{liu2025empirical}, overfitting fixes \cite{xu2025aligning,le2019automated}, reinventing or bypassing abstractions \cite{liu2025empirical}, and formatting or cosmetic noise \cite{campos2025empirical}, summarized in Table~\ref{tab:inflation-rules}.
(2) In rejection sampling, we use the issue description and code context from \textsc{SWE-bench Oracle} \cite{jimenez2024swe} to ask the LLM to write extensive patches from scratch, sampling 10 candidates per instance and keeping the largest.
With as-is examples, this yields 2,958 SFT instances.

\textbf{Preference-pair construction.}
For the 411 DPO seeds, we build larger input patches with the same procedures, then sample 10 refined patches from the SFT model per input.
Patches larger than the gold ones become rejected patches and the gold patches become chosen patches.
Since each seed contributes one inflated and one rejection-sampled input, this produces 822 DPO preference instances.


\subsection{Reasoning Trace Collection}
\label{sec:reasoning-trace-data}

Beyond input--output pairs, we collect reasoning traces via chain-of-thought (CoT) distillation \cite{wei2022chain}, where a high-performance closed-source LLM explains each refinement, transferring complex reasoning from stronger to smaller models \cite{zhuang2025unicott,zawalski2024robotic}.
We emphasize structured stepwise traces, which aid student-model learning and improve rationality \cite{zhuang2025unicott,li2025llms}.
Traces cover all three transformations, namely lower-CCR to higher-CCR solutions, tangled to atomic fix commits, and inflated to concise patches, with the teacher explaining how each concise target derives from the verbose patch \cite{yu2025cot,xu2026mind}.

\begin{table}
\centering
\caption{Overview of the patch refinement dataset.}
\label{tab:dataset-overview}
\setlength{\tabcolsep}{3pt}
\renewcommand{\arraystretch}{0.9} 
\scriptsize
\begin{tabular}{m{0.11\linewidth}|m{0.1\linewidth}|m{0.14\linewidth}|m{0.14\linewidth}|m{0.15\linewidth}|m{0.07\linewidth}|m{0.07\linewidth}}
\hline
\makecell[c]{Source} & \makecell[c]{Granu-\\larity} & \makecell[c]{Input\\patch} & \makecell[c]{Target\\patch} & \makecell[c]{Relevant\\context} & \makecell[c]{Train} & \makecell[c]{\#\\Inst.} \\
\hline
ACPR & Function & \makecell[c]{Lower-\\CCR\\solution} & \makecell[c]{Higher-\\CCR\\solution} & \makecell[c]{Problem\\description} & SFT & 1,227 \\
\hline
\makecell[l]{CCS\\commits} & Commit & \makecell[c]{Tangled\\fix commit} & \makecell[c]{Atomic\\fix commit} & \makecell[c]{Fix\\message} & SFT & 533 \\
\hline
\multirow{8}{*}{\makecell[l]{Synthetic}} & Repo & \makecell[c]{Inflated / \\sampled\\patch} & Gold patch & \multirow{8}{*}{\makecell[c]{Issue\\description / \\repo\\context}} & SFT & 2,958 \\
 & Repo & \makecell[c]{Inflated / \\sampled\\patch} & \makecell[c]{Gold patch\\(chosen)\\SFT\\sampled\\(reject)} &  & DPO & 822 \\
\hline
\multicolumn{6}{l|}{Total} & 5,540 \\
\hline
\end{tabular}
\end{table}

\subsection{Final Dataset}
\label{sec:patch-format}

The final dataset contains 5,540 instances, namely 1,227 function-level pairs, 533 tangled-commit pairs, and 3,780 SWE-bench synthetic instances (Table~\ref{tab:dataset-overview}).
We use \textsc{SEARCH}/\textsc{REPLACE} (S/R) blocks for patches rather than Git diffs, avoiding the invalid or hallucinated line numbers of diff headers while improving application efficiency \cite{xia2024agentless}.
We recover the final Git diff deterministically from the applied blocks, so SWE-bench \cite{jimenez2024swe} evaluates line numbers from the diff algorithm rather than LLM-generated ones.

%% file: 4.approach.tex
\section{Approach}
\label{sec:approach}

\begin{figure*}[htbp]
\centerline{\includegraphics[width=.70\linewidth]{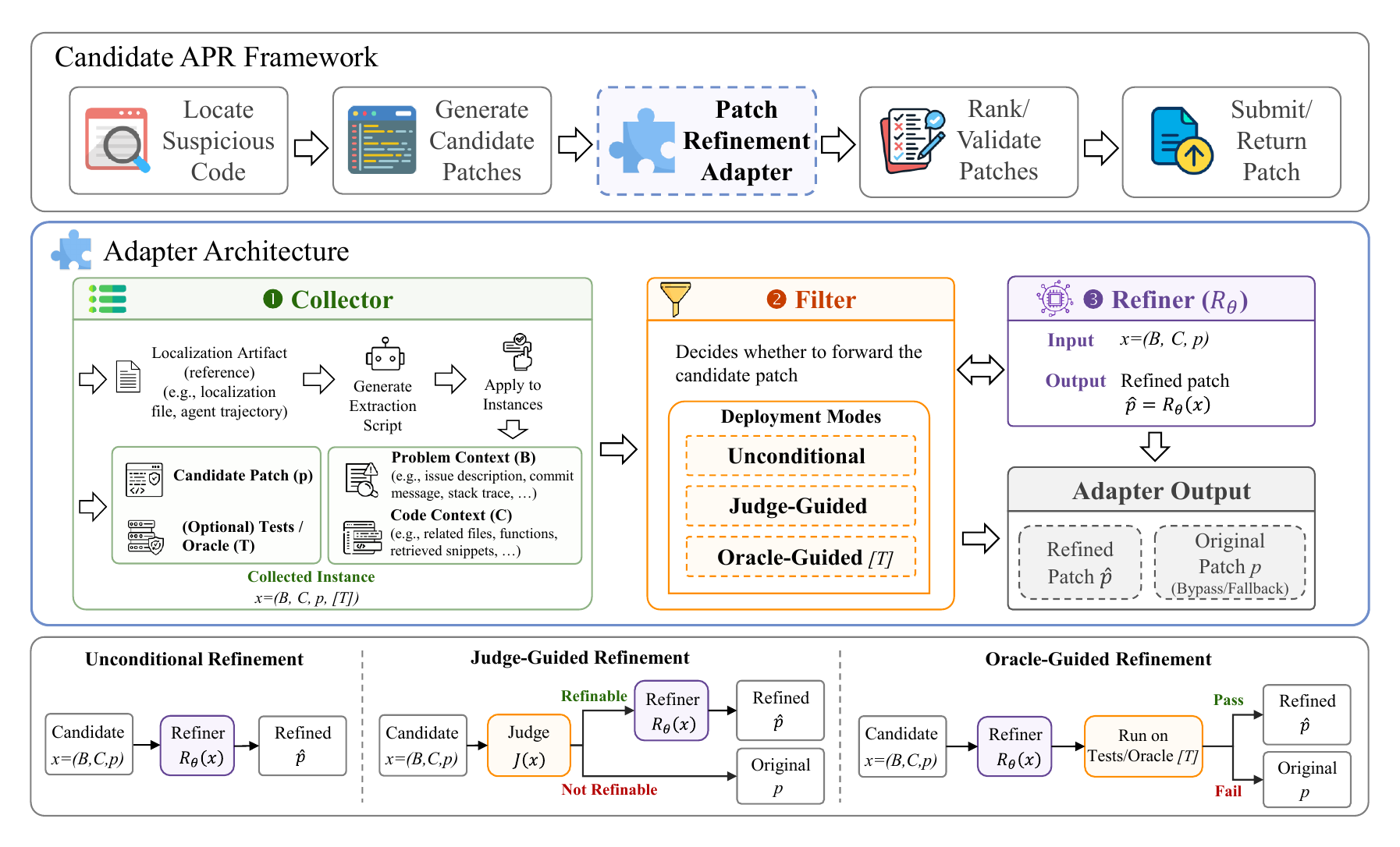}}
\caption{Overview of RECAP.}
\label{fig:overview}
\end{figure*}

APR is commonly a pipeline that locates suspicious code, generates candidate patches, and validates them against an oracle \cite{pereira2025systematic,yang2025patch,yang2025survey}.
Recent LLM-based systems instantiate it as fine-tuned models, prompt-based pipelines, multi-step workflows, or autonomous agents \cite{yang2025survey}, the generated patch remains the unit passed to validation, ranking, or submission.
This common interface motivates a refinement adapter that wraps existing frameworks.
As shown in Figure~\ref{fig:overview}, rather than replacing patch generation, we insert a lightweight plug-in refinement step between candidate generation and downstream stages.

\subsection{Task Formulation}
\label{sec:approach-formulation}

Let $B$ denote the problem context, $C$ the code context, and $p$ a candidate patch from the repair framework, where applying $p$ to program $P$ yields $P \oplus p$.
Given $x=(B,C,p)$, patch refinement produces a refined patch $p'$ preferable to $p$ under the intended bug-fixing specification.
The primary objective is concision, where among patches satisfying the intended behavior, $p'$ expresses the fix with a smaller, simpler change than $p$.
Correctness acts as a constraint, so $p'$ preserves a correct $p$ and may repair an incorrect $p$ when the context supports a better fix.
Formally, the correctness condition is:
\begin{equation}
  P \oplus p' \models \mathcal{S},
\end{equation}
where $\mathcal{S}$ is the intended bug-fixing specification.
This condition is conceptual since $\mathcal{S}$ is rarely available in real APR settings, and a test suite $T$ provides an executable oracle for plausibility rather than a proof of correctness.
Refinement is thus separated from generation, where the adapter starts from an existing candidate, improves its concision and logic, and must not make a correct or plausible candidate worse.

\subsection{Overview of RECAP}
\label{sec:adapter-overview}

As detailed in Figure~\ref{fig:overview}, RECAP has three components, namely a collector, a filter, and a refiner.
The \textbf{collector} standardizes the interface between a framework and RECAP.
Since frameworks expose artifacts in heterogeneous formats, it obtains context through a framework-adaptive procedure rather than a fixed parser.
From one reference localization artifact, e.g., a bug-localization file for workflow systems such as Agentless \cite{xia2024agentless} or a trajectory for agentic systems such as SWE-agent \cite{sweagent}, an LLM generates a reusable Python extraction script applied to all instances of that framework.
The candidate patch $p$ is located via configuration, and the optional test resource $T$ is gathered by a script run against the patched code, targeting SWE-bench suites but extensible to other oracles.
The collector emits a unified context file pairing each instance with $(B,C,p,T)$, giving the refiner a stable input format without assuming a single architecture or a hand-written parser per framework.

The \textbf{filter} decides whether a candidate is sent to the refiner according to the deployment mode (Section~\ref{sec:deployment-modes}), forwarding all candidates when refinement risk is acceptable or, in conservative modes, keeping the candidate unchanged or invoking an LLM-as-a-Judge to estimate whether refinement is safe.

The \textbf{refiner} is a small open-source LLM trained for patch refinement.
Given $x=(B,C,p)$, it generates $\hat{p}=R_\theta(x)$ in the same S/R representation as the input, and RECAP returns either $\hat{p}$ or the original $p$ according to the active mode.

\subsection{Training the Refiner}
\label{sec:training-refiner}

We train the refiner in two phases, where SFT teaches the input-output structure of patch refinement and DPO shifts its preference toward concise patches.
Both use the dataset constructed in Section~\ref{sec:data-construction}, randomly partitioned into training and evaluation splits at a 90\% / 10\% ratio.

\textbf{SFT phase.}
Let $\mathcal{D}_{\mathrm{sft}}=\{(x_i,y_i)\}_{i=1}^{N}$ be the supervised dataset, where $x_i=(B_i,C_i,p_i)$ and $y_i$ is the reasoning trace ending with the final S/R patch.
We fine-tune the refiner policy on $\mathcal{D}_{\mathrm{sft}}$ with a token-level cross-entropy loss over each target sequence $y_i$.

\textbf{Curriculum scheduling.}
The three sources form a natural difficulty progression by granularity, namely function-level pairs / tangled commits / synthetic repository-level instances (\emph{easy} / \emph{medium} / \emph{hard}).
We exploit this ordering with curriculum learning to stabilize SFT \cite{yang2025semantics}, with phase composition governed by a smoothing factor $\lambda$ \cite{yin2026improving}.
Training spans six epochs, starting with a warm-up on easy data only ($100/0/0$).
In each later epoch, $\lambda$ of the samples come from the hard source while the remaining $1-\lambda$ rehearse easy and medium sources to prevent catastrophic forgetting.
Raising $\lambda$ across the five epochs ($0,\,0.3,\,0.6,\,0.9,\,1.0$) yields hard ratios of $0/30/60/90/100\%$, so the model consolidates simpler refinements before specializing on repository-level patches.

\textbf{DPO phase.}
We apply DPO \cite{rafailov2023direct} to contrast concise targets with less-preferred model-sampled patches.
We use DPO over online RL such as PPO \cite{schulman2017proximal} and GRPO \cite{shao2024deepseekmath}, since optimizing a static preference dataset needs no separate reward model or on-policy rollouts, keeping preference tuning cheap and consistent with our plug-and-play design.

Let $\mathcal{D}_{\mathrm{dpo}}=\{(x_i,y_i^+,y_i^-)\}_{i=1}^{M}$ be the preference dataset, where $y_i^+$ is the concise gold patch and $y_i^-$ is sampled from the SFT model (Section~\ref{sec:synthetic-data}). Taking the SFT model as the reference policy, we optimize the DPO objective so the refiner assigns higher relative likelihood to $y_i^+$ than to $y_i^-$, and the resulting policy serves as RECAP's refiner.

\subsection{Deployment Modes}
\label{sec:deployment-modes}

APR frameworks differ in the validation support available at refinement time, from running tests before accepting a patch to delayed validation, manual review, or benchmark infrastructure.
To match these settings, RECAP supports three modes (bottom of Figure~\ref{fig:overview}).

\textbf{Unconditional refinement (UR).}
The default mode, where the filter forwards every candidate patch to the refiner:
\begin{equation}
  A_{\mathrm{UR}}(x) = R_\theta(x).
\end{equation}
It maximizes concision opportunities and suits hosts that independently validate or rank all candidates downstream, at the risk that an imperfect refinement may alter a candidate that would otherwise be accepted.

\textbf{Judge-guided refinement (JGR).}
When tests are unavailable or too costly after refinement, the filter invokes an LLM-as-Judge $J$ to decide refinability via a three-step policy.
(1) An apparently incorrect patch is refinable since the refiner may repair it, (2) a correct but already minimal patch is unchanged, and (3) a correct but verbose patch is refined only when the judge deems it low-risk to correctness.
It then applies:
\begin{equation}
  A_{\mathrm{JGR}}(x) =
  \begin{cases}
    R_\theta(x), & \text{if } J(x)=\textsc{refinable},\\
    p, & \text{otherwise.}
  \end{cases}
\end{equation}
This trades some refinement opportunity for a conservative fallback, avoiding changes to patches that are already minimal or whose verbosity may be required for correctness.

\textbf{Oracle-guided refinement (OGR).}
When an executable oracle $T$ is available, RECAP refines every candidate, applies it, and runs the tests, returning the refined patch only when the patched program passes and otherwise falling back to the original:
\begin{equation}
  A_{\mathrm{OGR}}(x,T) =
  \begin{cases}
    \hat{p}, & \text{if } P \oplus \hat{p} \models T,\\
    p, & \text{otherwise,}
  \end{cases}
  \quad \text{where } \hat{p}=R_\theta(x).
\end{equation}
This is the strongest safeguard, though bounded by test-suite strength, since a patch passing $T$ is only plausible with respect to that oracle.

%% file: 5.evaluation.tex
\section{Evaluation}
\label{sec:evaluation}


\subsection{Research Questions}

\textbf{RQ-1: To what extent do SOTA APR approaches produce verbose patches, and what factors contribute to verbosity?}
RQ-1 motivates patch refinement by quantifying verbosity in current APR approaches and analyzing associated design factors. It has two sub-questions.

\textbf{RQ-1.1}: \emph{\textbf{What is the status quo of patches generated by SOTA APR approaches?}}
This question measures whether current high-performing systems tend to produce patches larger and more verbose than developer patches.

\textbf{RQ-1.2}: \emph{\textbf{What causes different patch-size outcomes?}}
This question examines whether observable design choices are associated with different patch-size patterns.

\textbf{RQ-2: How effective is RECAP in refining patch size while maintaining correctness?}
RQ-2 evaluates whether RECAP can reduce patch size, scope, and complexity on patches produced by SOTA approaches while preserving the correctness of resolved SWE-bench instances.

\textbf{RQ-3: How do different training components contribute to patch refinement?}
RQ-3 ablates the SFT and DPO phases of Section~\ref{sec:training-refiner} to show how each affects concision and correctness preservation.

\subsection{Experimental Setup}

\subsubsection{Model}
We use \textbf{GPT-5.2} \cite{gpt52} for concise-target judging, patch inflation, and rejection sampling, since noisy synthetic labels would degrade the refinement supervision, and \textbf{Gemini-3-Flash} \cite{gemini3} for reasoning-trace collection, as its traces are more consistently organized as stepwise procedures.
Our main refiner is \textbf{QWen-3.5-27B} \cite{qwen3.5}, a recent small model, since the refiner must stay compact while handling long-context repository-level input.
For RQ-2, we use \textbf{Claude-4.5-Sonnet} \cite{cs45} as the pure-prompting baseline to test whether a small specialized refiner can outperform a stronger general-purpose model, and we also instantiate RECAP with \textbf{Gemma-4-E4B-it} \cite{gemma4} and \textbf{Mistral-3-14B-Reasoning} \cite{mistral3} for generalizability.
All selected models have at least a 64K-token context window to support repository-level SWE-bench instances.

\subsubsection{Benchmark}
We evaluate on SWE-bench Verified \cite{jimenez2024swe}, the human-filtered subset of SWE-bench.
Its instances are complex real-world repository-level GitHub issues requiring broad code context and multi-file reasoning, and it is a practical evaluation standard for LLM-based APR agents, enabling direct comparison with recent SOTA systems.

\subsubsection{Metrics}
We measure patch size from three complementary perspectives, namely textual size, edit scope, and code complexity.
For textual size, we follow prior LOC-based measurements of code change and review effort \cite{zhang2015interactive,shariffdeen2020automated,thongtanunam2017review,baysal2016investigating}, excluding context lines since these depend on diff configuration (Git defaults to three \cite{gitdiff}).

\textbf{Total Changes}. The total number of the added LOC and the deleted LOC (added + deleted).

\begin{table*}[t]
\centering
\caption{Comparison between agent-generated patches and corresponding gold patches on SWE-bench Verified.}
\label{rq1_agents}

\setlength{\tabcolsep}{1pt} %
\renewcommand{\arraystretch}{1.1}

\resizebox{\textwidth}{!}{%
\begin{tabular}{l c r r r r r r r | l c r r r r r r r}
\hline
\textbf{\makecell{Approach}} & \textbf{\makecell{LLM}} & \textbf{\makecell{\# Re-\\solved}} & \textbf{\makecell{$\Delta$Total\\Changes \%}} & \textbf{\makecell{$\Delta$Net\\Changes \%}} & \textbf{\makecell{$\Delta$Edited\\Files \%}} & \textbf{\makecell{$\Delta$Halstead\\Time Diff\%}} & \textbf{\makecell{$\Delta$CC\\Diff\%}} & \textbf{\makecell{\# w/\\aux edits}} & \textbf{Approach} & \textbf{\makecell{LLM}} & \textbf{\makecell{\# Re-\\solved}} & \textbf{\makecell{$\Delta$Total\\Changes \%}} & \textbf{\makecell{$\Delta$Net\\Changes \%}} & \textbf{\makecell{$\Delta$Edited\\Files \%}} & \textbf{\makecell{$\Delta$Halstead\\Time Diff\%}} & \textbf{\makecell{$\Delta$CC\\Diff \%}} & \textbf{\makecell{\# w/\\aux edits}} \\ 
\hline
Trae \cite{gao2025trae}              & Multiple          & 376                                  & +97.85                                          & +58.64                                        & +3.33                                          & +275.97                                             & +18.04                                   & 0& Swe-Rizzo \cite{swerizzo}         & Claude 3.7 Sonnet & 283                                  & +30.38                                          & -6.31                                         & -0.80                                          & +62.4                                               & -0.83                                     & 1                                        \\
Lingxi \cite{yang2025lingxi}            & Claude 4 Sonnet   & 373                                  & +81.59                                          & +59.53                                        & +2.89                                          & +72.17                                              & +43.15                                   & 0                                       & CodeSweep \cite{CodeSweep}         & Kimi K2 Instruct  & 267                                  & +226.73                                         & +178.71                                       & +14.45                                         & -102.73                                             & +73.28                                    & 8                                       \\
JoyCode \cite{JoyCode}           & Multiple          & 373                                  & +114.37                                         & +59.39                                        & +11.66                                         & +192.05                                             & +50.81                                   & 40& Agentless \cite{xia2024agentless}         & Claude 3.5 Sonnet & 254                                  & +17.62                                          & -16.25                                        & -2.61                                          & +15.92                                              & -22.57                                    & 0                                        \\
Refact \cite{Refact}            & Claude 4 Sonnet   & 372                                  & +654.31                                         & +572.15                                       & +63.33                                         & +533.95                                             & +164.14                                  & 200& Composio \cite{composio}          & Multiple          & 243                                  & +88.94                                          & +20.84                                        & +2.47                                          & -5.5                                                & -63.11                                    & 5                                       \\
OpenHands \cite{wang2025openhands}         & GPT 5             & 359                                  & +126.6                                          & +114.01                                       & +10.24                                         & +170.33                                             & +57.18                                   & 6& Navie \cite{navie}             & GPT 4o            & 236                                  & +110.57                                         & +55.4                                         & -1.69                                          & -116.15                                             & +32.04                                    & 1                                        \\
Moatless \cite{moatless}          & Claude 4 Sonnet   & 354                                  & +133.44                                         & +136.76                                       & +8.04                                          & +67.4                                               & +71.2                                    & 4& Skywork \cite{zeng2025skywork}           & Skywork-SWE-32B   & 235                                  & +227.43                                         & +67.43                                        & +41.70                                         & -4.51                                               & +98.32                                    & 42                                       \\
Augment \cite{augment}           & Claude 4 Sonnet   & 352                                  & +192.19                                         & +100.47                                       & +14.92                                         & +152.17                                             & +57.1                                    & 38& AutoCodeRover \cite{zhang2024autocoderover}     & Claude 3.5 Sonnet & 231                                  & +20.75                                          & -10.47                                        & -0.72                                          & -112.18                                             & -4.17                                     & 0                                        \\
Zai \cite{Zai}               & GLM 4.6           & 341                                  & +15143.45                                       & +15220.49                                     & +505.06                                        & -14812.27                                           & +2829.4                                  & 320& Gru \cite{gru}               & -                 & 226                                  & +33.83                                          & +0.31                                         & +7.23                                          & +34.05                                              & -4.81                                     & 1                                        \\
Cortexa \cite{cortexa}           & GPT o3            & 341                                  & +118.22                                         & +104.37                                       & -0.88                                          & -29.17                                              & +25.87                                   & 1                                       & SWE-Exp \cite{sweexp}           & DeepSeek V3       & 210                                  & +64.70                                          & +9.83                                         & +0.73                                          & -1936.89                                            & +14.47                                    & 1                                        \\
SWE-agent \cite{sweagent}          & Claude 4 Sonnet   & 333                                  & +690.91                                         & +1158.45                                      & +37.83                                         & +216.61                                             & +116.67                                  & 43& SWE-agent \cite{sweagent}          & SWE-agent-LM-32B  & 201                                  & +4461.9                                         & +5207.17                                      & +297.51                                        & -313.78                                             & +493.42                                   & 201                                      \\
PatchPilot \cite{li2025patchpilot}        & GPT o4 mini       & 323                                  & +125.33                                         & +94.38                                        & +17.78                                         & -23.16                                              & +10.68                                   & 2                                       & SWE-agent \cite{sweagent}          & Claude 3.5 Sonnet & 168                                  & +1009.05                                        & +242.23                                       & +134.52                                        & -87.23                                              & +136.32                                   & 136                                      \\
SWE-agent \cite{sweagent}          & Claude 3.7 Sonnet & 312                                  & +275.87                                         & +216.76                                       & +19.10                                         & +266.76                                             & +83.62                                   & 10& SWE-Fixer \cite{swefixer}          & QWen 2.5  & 164                                  & +69.14                                          & +17.69                                        & +34.45                                         & +23.77                                              & -26.09                                    & 0                                        \\
EntroPO \cite{entropo}           & QWenCoder 30B     & 302                                  & +4503.46                                        & +5195.13                                      & +191.11                                        & +901.14                                             & +638.32                                  & 239                                     & Lingma \cite{lingma}            & SWE-GPT-72B       & 144                                  & +61.41                                          & +8.12                                         & +5.23                                          & -800.61                                             & +4.3                                      & 8                                       \\
DeepSWE \cite{deepswe2025}         & DeepSWE           & 294                                  & +69.95                                          & +21.07                                        & +18.62                                         & -333.14                                             & +25.49                                   & 57                                      & SWE-agent \cite{sweagent}          & GPT 4o            & 116                                  & +465.59                                         & +759.37                                       & +90.52                                         & -396.82                                             & +44.83                                    & 33                                      \\ 
\hline
-                 & -                 & -                                    & -                                               & -                                             & -                                              & -                                                   & -                                        & -                                       & MEDIAN            & -                 & 288.50                               & +121.78                                         & +80.91                                        & +13.06                                         & +5.71                                               & +43.99                                    & 7                                    \\
\hline
\end{tabular}%
}
\end{table*}

\textbf{Net Changes}. The difference between added and deleted LOC (added - deleted).

\textbf{Edited Files}. Count of files modified by a patch, capturing edit scope as verbose patches may touch unrelated files.
We also measure how a patch changes the complexity of the modified source files.

\textbf{Halstead Time} \cite{halstead1977elements}. An estimated implementation time derived from Halstead effort, computed from the counts of distinct and total \emph{operators} and \emph{operands}, a more interpretable member of the Halstead family than effort alone.

\textbf{Cyclomatic Complexity (CC)} \cite{ebert2016cyclomatic}. The number of linearly independent paths in the control-flow graph, capturing whether a patch introduces additional branching structure.

Since Halstead and CC are computed on complete source files, for each patch we compute the metrics on the original files, apply the patch, recompute on the modified files, and treat the difference as the introduced complexity change.

Absolute sizes can mislead, since harder instances may need larger patches.
Following the relative-comparison principle in prior repair evaluation \cite{gulwani2018automated}, we compute the per-instance ratio between each generated patch and its gold patch, average across instances, and subtract 100\% to obtain the $\Delta$ values, where positive values indicate larger or more complex changes than gold and negative the opposite.

\subsection{RQ-1 (Patch Verbosity in SOTA APR)}

\textbf{[Experimental Design for RQ-1]:}
From the SWE-bench Verified leaderboard \cite{swebenchlead} (November 2025), we collect open- and closed-source submissions throughout the list, keeping the latest version per approach except SWE-agent \cite{sweagent} where we retain all LLM variants to reveal how the underlying model affects patch size, yielding 28 approaches.

For RQ-1.1, we average $\Delta$ values over each approach's resolved instances to measure successful-patch size relative to gold.
Since some patches also modify tests harness, or reproduction scripts that inflate LOC for non-fixing purposes, we manually check each patch and report resolved instances containing auxiliary test or reproduction edits (column \textit{\# w/ aux edits}).

For RQ-1.2, we inspect papers, technical reports, and repositories to classify each approach along five dimensions:
\textbf{(1) patch format} (structured formats such as S/R, XML, or JSON versus Git diff),
\textbf{(2) minimality instruction} (explicit, implicit, or absent size-constraining instructions),
\textbf{(3) context scope} (fixed window, structured such as AST-level, full repository, or dynamic selection),
\textbf{(4) iterative refinement} (whether a candidate is revised iteratively), and
\textbf{(5) resolution performance} (top- versus bottom-ranked by resolved instances).
Focusing on Total and Net Changes, we run instance-wise tests per group, using the Mann-Whitney-Wilcoxon (MWW) test \cite{mann1947test,wilcoxon1992individual} for binary dimensions and the Kruskal-Wallis H test \cite{ostertagova2014methodology} for multi-group ones, with pairwise MWW tests under Holm--Bonferroni correction \cite{holm1979simple} when significant and Cliff's Delta \cite{cliff1993dominance} for effect magnitude.

\textbf{[Experimental Results for RQ-1.1]:} \textit{\textbf{Verbosity is systemic, not an outlier effect.}} Table~\ref{rq1_agents} shows successful patches are consistently larger than gold. All 28 approaches exceed gold with the median producing +121.78\% more total changes, +80.91\% more net changes, and +13.06\% more edited files. Since every approach rather than a few outliers shows inflation, verbosity is a systemic property of current APR systems, not only with particular frameworks or models. Even successful repairs broaden the change footprint a reviewer must inspect.

\textit{\textbf{Textual size and operator-level complexity can move in opposite directions.}} While LOC inflation is uniformly positive, the Halstead Time difference is widely dispersed, from $-14812\%$ (Zai) to $+901\%$ (EntroPO), with a near-zero median of $+5.71\%$ that reflects cancellation rather than a small increase. Two factors explain the divergence from the LOC trend. First, Total Changes count both additions and deletions, so a textually large patch can leave the final file's operator and operand volumes unchanged or even simpler. Second, part of the LOC inflation comes from auxiliary edits to tests and reproduction scripts rather than production logic. These edits are common but uneven, ranging from 0 to 320 resolved instances per approach (median 7) and clustering in the most verbose systems. The four approaches exceeding $+1000\%$ total changes are exactly the four most aux-heavy ones, namely Zai ($+15143\%$, 320), EntroPO ($+4503\%$, 239), SWE-agent with SWE-agent-LM-32B ($+4462\%$, 201), and SWE-agent with Claude 3.5 Sonnet ($+1009\%$, 136). For these systems raw patch size overstates the production change a reviewer must reason about, confirming textual size alone is unreliable to interpret the patch complexity.

\begin{table}[t]
\centering
\caption{Significance test results for patch format, iterative refinement, and resolution.}
\label{rq1_p_1}
\scriptsize
\setlength{\tabcolsep}{4pt}
\renewcommand{\arraystretch}{0.9}
\begin{tabular}{c | c c | c}
\hline
\multirow{2}{*}{\textbf{Patch Format}} & \multirow{2}{*}{\textbf{Structured}} & \multirow{2}{*}{\textbf{Git-Diff}} & \textbf{p, $|\delta|$} \\
 & & & \textbf{S vs. G} \\ 
\hline
$\Delta$Avg Total Changes \% & +130.29 & +227.31 & 0.141, 0.02 \\
$\Delta$Avg Net Changes \% & +91.93 & +243.39 & \textbf{0.042}, 0.03 \\ 
\hline
\multirow{2}{*}{\textbf{Iterative Refinement}} & \multirow{2}{*}{\textbf{w/}} & \multirow{2}{*}{\textbf{w/o}} & \textbf{p, $|\delta|$} \\
 & & & \textbf{w/ vs. w/o} \\ 
\hline
$\Delta$Avg Total Changes \% & +406.46 & +95.60 & \textbf{\textless{}0.001}, 0.21 \\
$\Delta$Avg Net Changes \% & +496.64 & +47.09 & \textbf{\textless{}0.001}, 0.17 \\ 
\hline
\multirow{2}{*}{\textbf{Resolution}} & \multirow{2}{*}{\textbf{Top-12}} & \multirow{2}{*}{\textbf{Bottom-12}} & \textbf{p, $|\delta|$} \\
 & & & \textbf{T vs. B} \\ 
\hline
$\Delta$Avg Total Changes \% & +223.39 & +118.09 & \textbf{\textless{}0.001}, 0.17 \\
$\Delta$Avg Net Changes \% & +224.67 & +90.39 & \textbf{\textless{}0.001}, 0.19 \\
\hline
\end{tabular}
\end{table}

\textit{\textbf{Larger patches are also structurally heavier, not merely longer.}} The median Cyclomatic Complexity difference is $+43.99\%$, almost eight times the Halstead Time median, showing verbosity is not driven solely by bulk additions, deletions, or auxiliary edits, since patches also add branching structure to production code, a distinct review and maintenance concern. Current LLM-based APR agents thus produce successful patches that are simultaneously larger, broader in file scope, and structurally more complex than gold, motivating a downstream refinement step that targets both edit volume and structural complexity rather than LOC alone.

\findx{\textbf{[RQ-1.1] Finding:}
\textit{\textbf{Patch verbosity is widespread and multi-dimensional.}}
All 28 evaluated SWE-bench approaches exceed the gold patch in total changes (median +121.78\%), net changes (+80.91\%), edited files (+13.06\%), and cyclomatic complexity (+43.99\%), indicating that verbosity is prevalent across current systems. Textual size and structural complexity can move in opposite directions, demonstrating that patch verbosity cannot be captured by single one size metric.}

\textbf{[Experimental Results for RQ-1.2]:} 
\textit{\textbf{Surface-level controls are weak.}} Tables~\ref{rq1_p_1} and~\ref{rq1_p_2} relate patch size to five design choices. Output format shows limited association. Structured formats have lower average total changes than Git-diff (+130.29\% vs. +227.31\%) but not significantly so, while net changes are significant (+91.93\% vs. +243.39\%) with a negligible effect ($|\delta|=0.03$). Minimality instructions are more striking, since explicitly instructing minimality (+208.83\% total, +168.05\% net) yields patches as large as no instruction (+191.39\%, +194.09\%), while implicit guidance gives the smallest patches (+49.65\%, +9.56\%), suggesting that the underlying repair mechanism dictates patch size far more than the prompt wording. Both dimensions indicate that formatting or instructing toward minimality, without changing how the repair is produced, barely affects verbosity.

\begin{table}[htbp]
\centering
\caption{Significance test results for minimality instruction, and context scope.}
\label{rq1_p_2}
\scriptsize
\setlength{\tabcolsep}{4pt}
\renewcommand{\arraystretch}{0.90}
\begin{tabular}{ll cc}
\hline
 & \textbf{Pair} & \makecell{\textbf{$\Delta$Avg Total}\\\textbf{Changes \%}} & \makecell{\textbf{$\Delta$Avg Net Code}\\\textbf{Size Change \%}} \\
\hline
\multicolumn{4}{c}{\textbf{Minimality Instruction}} \\
\hline
\multirow{3}{*}{} & Explicit (E) & +208.83 & +168.05 \\
 & Implicit (I) & +49.65 & +9.56 \\
 & None (N) & +191.39 & +194.09 \\
\hline
KW Test & $p$  & \textbf{\textless{}0.001} & \textbf{\textless{}0.001} \\
\hline
\multirow{3}{*}{\makecell[l]{Pairwise\\$p, |\delta|$}} & N vs. E & \textbf{\textless{}0.001}, 0.15 & \textbf{\textless{}0.001}, 0.14 \\
 & N vs. I & \textbf{\textless{}0.001}, 0.13 & \textbf{\textless{}0.001}, 0.15 \\
 & E vs. I & \textbf{\textless{}0.001}, 0.28 & \textbf{\textless{}0.001}, 0.29 \\
\hline
\multicolumn{4}{c}{\textbf{Context Scope}} \\
\hline
\multirow{4}{*}{} & Fixed Window (FW) & +300.34 & +398.57 \\
 & Structured (S) & +72.93 & +43.12 \\
 & Full Repo (FR) & +221.23 & +139.70 \\
 & Dynamic (D) & +105.99 & +62.22 \\
\hline
KW Test & $p$ & \textbf{\textless{}0.001} & \textbf{0.005} \\
\hline
\multirow{6}{*}{\makecell[l]{Pairwise\\$p, |\delta|$}} & FW vs. S & \textbf{\textless{}0.001}, 0.12 & \textbf{0.02}, 0.06 \\
 & FW vs. FR & 0.23, 0.03 & 0.71, 0.01 \\
 & FW vs. D & 0.23, 0.03 & 0.52, 0.03 \\
 & S vs. FR & \textbf{\textless{}0.001}, 0.14 & \textbf{0.01}, 0.07 \\
 & S vs. D & \textbf{\textless{}0.001}, 0.09 & 0.52, 0.04 \\
 & FR vs. D & \textbf{0.01}, 0.06 & 0.52, 0.04 \\
\hline
\end{tabular}
\end{table}

\textit{\textbf{Capability-oriented design enlarges patches.}} The strongest, most consistent trend is that higher-capability design also means larger patches. Iterative refinement produces far larger patches than its absence (+406.46\% vs. +95.60\% total, +496.64\% vs. +47.09\% net), and top-ranked systems exceed bottom-ranked ones (+223.39\% vs. +118.09\% total, +224.67\% vs. +90.39\% net), all four contrasts significant. The mechanisms that raise resolution also inflate the patch, a tension prompting cannot resolve, which is why we position refinement downstream to reduce verbosity after a capable agent finds a plausible repair without weakening the generator.

\begin{table*}[t]
\centering
\caption{Performance comparison between different baselines and RECAP on SWE-bench Verified.}
\label{rq2_cmp}

\setlength{\tabcolsep}{1pt}
\renewcommand{\arraystretch}{0.95}
\scriptsize

\begin{tabular}[t]{l|r|rrr|rr|} 
\hline
\makecell{Approach} & \makecell{Resolved} & \makecell{$\Delta$Total\\Chg \%} & \makecell{$\Delta$Net\\Chg \%} & \makecell{$\Delta$Files\\Aff \%} & \makecell{$\Delta$Halstead\\Time Diff \%} & \makecell{$\Delta$CC\\Diff \%} \\
\hline
Agentless & 254/500 & +17.62 & -16.25 & -2.61 & +15.92 & -22.57 \\
Prompting & 215/500\textcolor[rgb]{0.753,0,0}{(-39)} & +1.02 & -36.41 & -2.42 & +4.93 & -23.53 \\
Atomizer & 94/500\textcolor[rgb]{0.753,0,0}{(-160)} & +8.73 & +4.54 & -1.56 & \textbf{-20.57} & -16.32 \\
Adapatcher & 204/500\textcolor[rgb]{0.753,0,0}{(-50)} & +17.48 & -20.49 & -2.66 & +18.24 & -21.20 \\
PRepair & 205/500\textcolor[rgb]{0.753,0,0}{(-49)} & +18.81 & -12.85 & -2.77 & +22.90 & -20.13 \\
\rowcolor[rgb]{0.906,0.902,0.902} RECAP-UR & 258/500\textcolor[rgb]{0.082,0.502,0.239}{(+4)} & \textbf{-9.94} & \textbf{-50.40} & \underline{-2.86} & \underline{-16.50} & \underline{-23.79} \\
\rowcolor[rgb]{0.906,0.902,0.902} RECAP-JGR & \underline{265}/500\textcolor[rgb]{0.082,0.502,0.239}{(+11)} & -7.61 & \underline{-50.09} & -2.76 & -6.67 & -\textbf{24.09} \\
\rowcolor[rgb]{0.906,0.902,0.902} RECAP-OGR & \textbf{280}/500\textcolor[rgb]{0.082,0.502,0.239}{(+26)} & \underline{-7.64} & -47.71 & \textbf{-2.90} & -12.11 & -22.40 \\
\hline
SWE-agent & 333/500 & +690.91 & +1158.45 & +37.83 & +216.61 & +116.67 \\
Prompting & 303/500\textcolor[rgb]{0.725,0.11,0.11}{(-30)} & +41.70 & +9.55 & \underline{-0.30}& \textbf{-27.89} & +0.32 \\
Atomizer & 116/500\textcolor[rgb]{0.753,0,0}{(-217)} & +933.35 & +1064.78 & +23.4 & +377.00 & +202.04 \\
Adapatcher & 271/500\textcolor[rgb]{0.753,0,0}{(-62)} & +221.00 & +266.03 & +15.31 & +93.55 & +132.00 \\
PRepair & 255/500\textcolor[rgb]{0.753,0,0}{(-78)} & +461.96 & +275.33 & +5.48 & \underline{-1.00} & \textbf{-178.9} \\
\rowcolor[rgb]{0.906,0.902,0.902} RECAP-UR & \underline{357}/500\textcolor[rgb]{0.082,0.502,0.239}{(+24)} & \textbf{+24.69} & \textbf{-33.17} & \textbf{-1.93} & +30.89 & \underline{-12.48} \\
\rowcolor[rgb]{0.906,0.902,0.902} RECAP-JGR & 355/500\textcolor[rgb]{0.082,0.502,0.239}{(+22)}& \underline{+31.52}& \underline{-26.56}& +0.88& +24.16& -9.77\\
\rowcolor[rgb]{0.906,0.902,0.902} RECAP-OGR & \textbf{375}/500\textcolor[rgb]{0.082,0.502,0.239}{(+42)} & +52.39 & +50.64 & +3.40 & +65.30 & +18.80 \\
\hline
\end{tabular}%
\begin{tabular}[t]{l|r|rrr|rr}
\hline
\makecell{Approach} & \makecell{Resolved} & \makecell{$\Delta$Total\\Chg \%} & \makecell{$\Delta$Net\\Chg \%} & \makecell{$\Delta$Files\\Aff \%} & \makecell{$\Delta$Halstead\\Time Diff \%} & \makecell{$\Delta$CC\\Diff \%} \\
\hline
Moatless & 354/500 & +133.44 & +136.76 & +8.04 & +67.40 & +71.20 \\
Prompting & 310/500\textcolor[rgb]{0.725,0.11,0.11}{(-44)} & +22.34 & +28.72 & +1.22 & \textbf{-64.26} & +6.85 \\
Atomizer & 152/500\textcolor[rgb]{0.753,0,0}{(-202)} & +123.48 & +183.29 & +20.64 & +43.87 & +107.32 \\
Adapatcher & 265/500\textcolor[rgb]{0.753,0,0}{(-89)} & +85.20 & +61.90 & \underline{-1.65} & +55.15 & +37.74 \\
PRepair & 252/500\textcolor[rgb]{0.753,0,0}{(-102)} & +67.29 & +60.47 & +5.40 & +7.25 & +31.94 \\
\rowcolor[rgb]{0.906,0.902,0.902} RECAP-UR & \underline{350}/500\textcolor[rgb]{0.725,0.11,0.11}{(-4)} & \textbf{+5.09} & \textbf{-37.38} & \textbf{-1.99} & -18.73 & \textbf{-14.49} \\
\rowcolor[rgb]{0.906,0.902,0.902} RECAP-JGR & 348/500\textcolor[rgb]{0.725,0.11,0.11}{(-6)} & \underline{+10.32} & -29.88 & -1.57 & \underline{-28.92} & \underline{-6.91} \\
\rowcolor[rgb]{0.906,0.902,0.902} RECAP-OGR & \textbf{378}/500\textcolor[rgb]{0.082,0.502,0.239}{(+24)} & +14.10 & \underline{-32.76} & -0.90 & -2.42 & -5.58 \\
\hline
Openhands & 359/500 & +126.60 & +114.01 & +10.24 & +170.33 & +57.18 \\
Prompting & 330/500\textcolor[rgb]{0.725,0.11,0.11}{(-29)} & +7.74 & -15.08 & \underline{-1.97} & \textbf{-21.10} & \underline{-9.31} \\
Atomizer & 164/500\textcolor[rgb]{0.753,0,0}{(-195)} & +109.43 & +74.71 & +8.18 & +335.79 & +79.87 \\
Adapatcher & 276/500\textcolor[rgb]{0.753,0,0}{(-83)} & +99.73 & +38.93 & +3.32 & +39.47 & +51.10 \\
PRepair & 264/500\textcolor[rgb]{0.753,0,0}{(-95)} & +74.11 & +25.02 & +0.25 & +41.41 & +14.70 \\
\rowcolor[rgb]{0.906,0.902,0.902} RECAP-UR & 363/500\textcolor[rgb]{0.082,0.502,0.239}{(+4)} & -\textbf{2.89} & \textbf{-38.06} & \textbf{-2.07} & +22.67 & \textbf{-12.99} \\
\rowcolor[rgb]{0.906,0.902,0.902} RECAP-JGR & \underline{365}/500\textcolor[rgb]{0.082,0.502,0.239}{(+6)} & \underline{+4.79} & \underline{-32.84} & -1.92 & +38.18 & -2.04 \\
\rowcolor[rgb]{0.906,0.902,0.902} RECAP-OGR & \textbf{382}/500\textcolor[rgb]{0.082,0.502,0.239}{(+23)} & +15.36 & +0.26 & 0 & \underline{-20.18} & -0.62 \\
\hline
\end{tabular}
\end{table*}

\textit{\textbf{Context should be organized, not just enlarged.}} More context is not better for concision. Structured context gives the smallest patches (+72.93\% total, +43.12\% net) and fixed-window the largest (+300.34\%, +398.57\%), with full-repository and dynamic in between. 
Structured context is significantly smaller than full-repository for both metrics and than dynamic for total changes, suggesting that more exposed code broadens edits while context curated around repair-relevant structure keeps changes localized.

\findx{\textbf{[RQ-1.2] Finding:}
\textit{\textbf{Verbosity is closely tied to repair capability rather than surface-level design choices.}}
Neither structured output formats nor explicit minimality prompts lead to meaningfully smaller patches. In contrast, iterative refinement (+406.46\% vs.\ +95.60\% total, $p<0.001$) and higher resolution (+223.39\% vs.\ +118.09\%, $p<0.001$) are consistently associated with larger patches, revealing a capability--concision tradeoff that motivates a downstream refinement more robust than constraining the generator.}

\subsection{RQ-2 (Patch Refinement Effectiveness)}

\textbf{[Experimental Design for RQ-2]:}
We compare the RECAP against four baselines.
The first is prompting Claude-4.5-Sonnet with the same input as RECAP (issue context, code context, and verbose candidate patch), a strong prompting-only alternative to a specialized refiner.
The second is Atomizer \cite{zhu2026atomizer}, a SOTA collaborative multi-agent commit-untangling technique, which we apply by treating a verbose patch as a tangled commit and separating the atomic fix.
The third is Adapatcher \cite{dai2025less}, a two-stage APR framework that seeks small fixes through adaptive repair.
The fourth is PRepair \cite{ke2026qimeng}, an edit-aware repair method using reward optimization for precise repair.
As minimality-aware generators rather than refiners, Adapatcher and PRepair receive the same context ($B$ and $C$) without the candidate patch $p$ and generate a concise patch directly, testing whether post-generation refinement matches precise-generation training.
We evaluate all three modes from Section~\ref{sec:deployment-modes}, namely UR, JGR, and OGR.

\textbf{[Experimental Results for RQ-2]:}
\textit{\textbf{A small specialized refiner dominates the size--correctness tradeoff.}}
Direct prompting with the substantially stronger Claude-4.5-Sonnet aggressively reduces size yet sacrifices 29 to 44 resolved instances per host, and the minimality-oriented baselines lose 49 to 217.
RECAP is the only method that reduces patch size while preserving or improving resolution in almost every setting.

\textit{\textbf{Existing minimality methods do not transfer to repository-level refinement.}}
The baseline failures are systematic.
Atomizer separates intents but never guarantees the extracted ``atomic'' fix is self-contained, so it frequently drops code needed to resolve the issue, explaining its catastrophic $-217$ on SWE-agent.
Adapatcher and PRepair, trained for function-level repair, see their locality assumptions break down on repository-level patches and degrade correctness across all four hosts.
This validates our design, since repository-level refinement is a distinct task that cannot be served by repurposing untangling or function-level minimization tools.

\textit{\textbf{Deployment modes form a controllable, partly counter-intuitive safety tradeoff.}}
UR and JGR maximize concision, cutting average total changes from $+242.14\%$ to $+4.24\%$ and $+9.76\%$ and net changes from $+348.24\%$ to $-39.75\%$ and $-34.84\%$.
OGR is intentionally less aggressive at $+18.55\%$ total and $-7.39\%$ net because it falls back whenever the refined patch fails the oracle, which guarantees it cannot drop an already-resolved instance yet still improves all four systems by $+23$ to $+42$.

\textit{\textbf{Refinement helps most for verbose agents but still benefits already-compact patches.}}
The gains scale with host verbosity.
For SWE-agent, with extremely large patches (+690.91\% total, +1158.45\% net), RECAP-UR collapses these to +24.69\% total and -33.17\% net while adding 24 resolved instances, and RECAP-JGR reaches +31.52\% and -26.56\% with +22.
For Moatless and Openhands, net changes drop from $+136.76\%$ and $+114.01\%$ to $-37.38\%$ and $-38.06\%$.
Even Agentless, whose patches are already near-gold, is pushed below the gold baseline in total and net changes, supporting refinement as a general-purpose downstream step.

\textit{\textbf{Concision is multi-dimensional, and textual reduction does not imply complexity reduction.}}
RECAP-UR reduces edited-file scope and Cyclomatic Complexity below the host on all four cases, but Halstead Time is mixed, falling for Agentless and Moatless while remaining positive for SWE-agent and Openhands.
This mirrors RQ-1.1, where fewer lines and files do not guarantee a proportional drop in operator--operand complexity, so refinement must be judged on multiple axes and a host-by-host complexity check should accompany any reported LOC reduction.

\findx{\textbf{[RQ-2] Finding:}
\textit{\textbf{RECAP achieves a substantially better size--correctness tradeoff.}}
While untangling and function-level minimization do not transfer to repository-level refinement, RECAP resolves up to 42 additional instances and reduces average patch size from +242.14\% to +4.24\% in total changes while preserving or improving correctness, demonstrating superior size–correctness tradeoff through task-specialized refinement.}

\subsection{RQ-3 (Ablation Study)}

\textbf{[Experimental Design for RQ-3]:}
Since the full adapter uses both SFT and DPO, we compare two variants, one removing DPO to use the SFT-only model, which tests whether supervised pairs alone teach concise repair, and another removing both phases to prompt the base instruction model with the same refinement input. 
We estimate how much behavior comes from the pretrained model and prompt format alone.
We run each under the RQ-2 setting with the same metrics, deploying UR by default.

\begin{table}
\centering
\caption{Performance comparison between RECAP with and without different components.}
\label{rq3_ab}
\setlength{\tabcolsep}{2pt}
\renewcommand{\arraystretch}{0.85}
\resizebox{\columnwidth}{!}{%
\begin{tabular}{l|r|rrr|rr}
\hline
\makecell{Approach} & \makecell{Resolved} & \makecell{$\Delta$Total\\Chg \%} & \makecell{$\Delta$Net\\Chg \%} & \makecell{$\Delta$Files\\Aff \%} & \makecell{$\Delta$Halstead\\Time Diff \%} & \makecell{$\Delta$CC\\Diff \%} \\
\hline
Agentless & 254/500 & +17.62 & -16.25 & -2.61 & +15.92 & -22.57 \\
w/o SFT\&DPO & \underline{260}/500\textcolor[rgb]{0.082,0.502,0.239}{(+6)} & +2.27 & -28.43 & \underline{-2.74} & +0.49 & -22.48 \\
w/o DPO & 255/500\textcolor[rgb]{0.082,0.502,0.239}{(+1)} & \underline{-3.21} & \underline{-35.83} & -2.67 & \textbf{-21.26} & \textbf{-25.68} \\
w/ SFT\&DPO & \textbf{258}/500\textcolor[rgb]{0.082,0.502,0.239}{(+4)} & \textbf{-9.94} & \textbf{-50.40} & \textbf{-2.86} & \underline{-16.50} & \underline{-23.79} \\
\hline
SWE-agent & 333/500 & +690.91 & +1158.45 & +37.83 & +216.61 & +116.67 \\
w/o SFT\&DPO & \underline{291}/500\textcolor[rgb]{0.725,0.11,0.11}{(-42)} & +62.43 & +60.15 & -0.66 & \underline{-59.99} & +28.80 \\
w/o DPO & 273/500\textcolor[rgb]{0.725,0.11,0.11}{(-60)} & \textbf{+11.26} & \underline{-16.90} & \textbf{-2.05} & \textbf{-118.13} & \underline{+9.10} \\
w/ SFT\&DPO & \textbf{357}/500\textcolor[rgb]{0.082,0.502,0.239}{(+24)} & \underline{+24.69} & \textbf{-33.17} & \underline{-1.93} & +30.89 & \textbf{-12.48} \\
\hline
Moatless & 354/500 & +133.44 & +136.76 & +8.04 & +67.40 & +71.20 \\
w/o SFT\&DPO & \underline{304}/500\textcolor[rgb]{0.725,0.11,0.11}{(-50)} & +25.56 & +18.97 & -1.55 & -10.37 & +6.60 \\
w/o DPO & 280/500\textcolor[rgb]{0.725,0.11,0.11}{(-74)} & \underline{+8.32} & \underline{+6.72} & \textbf{-2.70} & \textbf{-35.30} & \underline{-9.60} \\
w/ SFT\&DPO & \textbf{350}/500\textcolor[rgb]{0.725,0.11,0.11}{(-4)} & \textbf{+5.09} & \textbf{-37.38} & \underline{-1.99} & \underline{-18.73} & \textbf{-14.49} \\
\hline
Openhands & 359/500 & +126.60 & +114.01 & +10.24 & +170.33 & +57.18 \\
w/o SFT\&DPO & \underline{317}/500\textcolor[rgb]{0.725,0.11,0.11}{(-42)} & +58.81 & -1.79 & -1.05 & +114.91 & \underline{-22.95} \\
w/o DPO & 295/500\textcolor[rgb]{0.725,0.11,0.11}{(-64)} & \underline{+44.46} & \underline{-3.15} & \underline{-1.38} & \textbf{-31.69} & \textbf{-24.90} \\
w/ SFT\&DPO & \textbf{363}/500\textcolor[rgb]{0.082,0.502,0.239}{(+4)} & \textbf{-2.89} & \textbf{-38.06} & \textbf{-2.07} & \underline{+22.67} & -12.99 \\
\hline
\end{tabular}
}
\end{table}

\textbf{[Experimental Results for RQ-3]:}
\textit{\textbf{Prompting alone refines patches but cannot be trusted with correctness.}}
Table~\ref{rq3_ab} shows the variant without SFT and DPO already shrinks patches relative to every host, confirming a general instruction model can perform shallow cleanup from the prompt alone.
This cleanup is unreliable, dropping 42 instances on SWE-agent, 50 on Moatless, and 42 on Openhands.

\textit{\textbf{SFT learns concision so well that it over-applies it, falling below even the untrained baseline.}}
Adding SFT sharpens concision across the board, lowering average total changes from +37.27\% to +15.21\%, average net changes from +12.23\% to -12.29\%, and average Halstead Time from +11.26\% to -51.60\% relative to the untrained variant, with the strongest Halstead reduction on all four hosts.
Yet supervised imitation reveals its blind spot, since the SFT-only model resolves only 276 instances on average, below both the full model (332) and the untrained variant (293).

\textit{\textbf{DPO supplies the missing restraint, and its payoff scales with verbosity.}}
DPO closes the gap by contrasting concise gold patches with the model's over-aggressive samples, injecting the negative signal SFT lacks.
The full model recovers average resolution from 276 to 332, concentrated where SFT over-cuts most, with +84 over SFT-only on SWE-agent, +70 on Moatless, and +68 on Openhands.
This recovery does not cost concision, since the full model still attains the lowest average total changes (+4.24\%), net changes (-39.75\%), and edited-file scope (-2.21\%) among all variants.

\textit{\textbf{Optimizing a single minimality metric is a trap.}}
SFT-only produces the best Halstead Time on every host and would look superior judged on that metric alone, yet it is the weakest model once correctness is considered.
The full model instead achieves better average Cyclomatic Complexity (-15.94\% vs. -12.77\%) with far higher resolution.
The lesson mirrors RQ-1, since the most aggressive simplifier is not the best patch, and the right target is a concise patch that still resolves the issue rather than the smallest syntactic or operator-level change.

\findx{\textbf{[RQ-3] Finding:}
\textit{\textbf{SFT teaches concision but over-applies it; DPO supplies the missing restraint that makes the refiner deployable.}}
The SFT-only model achieves the strongest size reduction (avg.\ total changes +15.21\%) but resolves fewer instances than even the untrained baseline (275.8 vs.\ 293.0), because it learns \emph{how} to compress but not \emph{when} compression breaks the fix.
Adding DPO recovers resolution to 332 (+84, +70, and +68 on the three most verbose hosts) without sacrificing concision, turning an aggressive simplifier into a usable adapter.}

%% file: 6.discussion.tex
\section{Discussion}

\begin{table}
\centering
\caption{Performance comparison between different LLMs on SWE-bench Verified.}
\label{llm_cmp}
\setlength{\tabcolsep}{2pt}
\renewcommand{\arraystretch}{0.85}
\resizebox{\columnwidth}{!}{%
\begin{tabular}{l|r|rrr|rr}
\hline
\makecell{Approach} & \makecell{Resolved} & \makecell{$\Delta$Total\\Chg \%} & \makecell{$\Delta$Net\\Chg \%} & \makecell{$\Delta$Files\\Aff \%} & \makecell{$\Delta$Halstead\\Time Diff \%} & \makecell{$\Delta$CC\\Diff \%} \\
\hline
Agentless     & 254/500 & +17.62 & -16.25 & -2.61 & +15.92 & -22.57 \\
Gemma 4 4B    & \textbf{273}/500\textcolor[rgb]{0.078,0.502,0.239}{(+19)} & \textbf{-11.37} & \underline{-40.95} & \underline{-2.61} & \underline{-15.83} & -21.78 \\
Mistral 3 14B & 255/500\textcolor[rgb]{0.078,0.502,0.239}{(+1)} & +0.72 & -19.58 & -2.58 & -1.23 & \textbf{-26.25} \\
Qwen 3.5 27B  & \underline{258}/500\textcolor[rgb]{0.078,0.502,0.239}{(+4)} & \underline{-9.94} & \textbf{-50.40} & \textbf{-2.86} & \textbf{-16.50} & \underline{-23.79} \\
\hline
SWE-agent     & 333/500 & +690.91 & +1158.45 & +37.83 & +216.61 & +116.67 \\
Gemma 4 4B    & 328/500\textcolor[rgb]{0.722,0.11,0.11}{(-5)} & +165.79 & +138.61 & 5.94 & 47.71 & 47.15 \\
Mistral 3 14B & \underline{337}/500\textcolor[rgb]{0.078,0.502,0.239}{(+4)} & \underline{+56.90} & \underline{+59.88} & \underline{3.05} & \textbf{-15.37} & \underline{28.15} \\
Qwen 3.5 27B  & \textbf{357}/500\textcolor[rgb]{0.078,0.502,0.239}{(+24)} & \textbf{+24.69} & \textbf{-33.17} & \textbf{-1.93} & \underline{+30.89} & \textbf{-12.48} \\
\hline
Moatless      & 354/500 & +133.44 & +136.76 & +8.04 & +67.40 & +71.20 \\
Gemma 4 4B    & 330/500\textcolor[rgb]{0.722,0.11,0.11}{(-24)} & \underline{+5.11} & \underline{-20.71} & \underline{-1.35} & \underline{-52.28} & 1.71 \\
Mistral 3 14B & \underline{346}/500\textcolor[rgb]{0.722,0.11,0.11}{(-8)} & +34.03 & +14.46 & -1.00 & \textbf{-54.53} & \underline{-8.26} \\
Qwen 3.5 27B  & \textbf{350}/500\textcolor[rgb]{0.722,0.11,0.11}{(-4)} & \textbf{+5.09} & \textbf{-37.38} & \textbf{-1.99} & -18.73 & \textbf{-14.49} \\
\hline
Openhands     & 359/500 & +126.60 & +114.01 & +10.24 & +170.33 & +57.18 \\
Gemma 4 4B    & 352/500\textcolor[rgb]{0.722,0.11,0.11}{(-7)} & \underline{22.84} & \underline{-21.26} & \textbf{-2.33} & +30.76 & \underline{+7.26} \\
Mistral 3 14B & \underline{360}/500\textcolor[rgb]{0.078,0.502,0.239}{(+1)} & 44.21 & 0.95 & -0.82 & \textbf{+7.14} & +16.05 \\
Qwen 3.5 27B  & \textbf{363}/500\textcolor[rgb]{0.078,0.502,0.239}{(+4)} & \textbf{-2.89} & \textbf{-38.06} & \underline{-2.07} & \underline{+22.67} & \textbf{-12.99} \\
\hline
\end{tabular}
}
\end{table}

\textbf{\textit{Model Generalizability.}}
A practical adapter should not be tied to one LLM backbone, we therefore re-instantiate it with two additional open-source models of different architectures and scales, Gemma-4-4B and Mistral-3-14B-Reasoning, evaluated under UR mode against the four RQ-2 hosts.
Table~\ref{llm_cmp} shows all three backbones reduce patch size on nearly every metric, indicating that the RQ-2 gains stem from the refinement task and training pipeline rather than a single backbone.
Even Gemma-4-4B is viable on all four hosts, enabled by modern long-context capacity that lets a 4B model ingest repository-level inputs exceeding the 32K-token windows of earlier small models.


\begin{table}[t]
\centering
\caption{Correctness transitions induced by refinement. C/I denote correct/incorrect
before$\rightarrow$after refinement.
Net $=$ Recovered $-$ Broken matches the resolution delta in RQ-2.}
\label{disc_transition}
\setlength{\tabcolsep}{4pt}
\renewcommand{\arraystretch}{0.85}
\footnotesize
\begin{tabular}{l|c|rrrr|r}
\hline
Host & Mode & C$\rightarrow$C & C$\rightarrow$I & I$\rightarrow$C & I$\rightarrow$I & Net \\
\hline
\multirow{3}{*}{Agentless}
 & UR  & 227 & 27 & 31 & 209 & \textcolor[rgb]{0.08,0.5,0.24}{+4}\\
 & JGR & 235 & 19 & 30 & 210 & \textcolor[rgb]{0.08,0.5,0.24}{+11}\\
 & OGR & 254 & 0  & 26 & 214 & \textcolor[rgb]{0.08,0.5,0.24}{+26}\\
\hline
\multirow{3}{*}{SWE-agent}
 & UR  & 314 & 19 & 43 & 124 & \textcolor[rgb]{0.08,0.5,0.24}{+24}\\
 & JGR & 318 & 15 & 37 & 130 & \textcolor[rgb]{0.08,0.5,0.24}{+22}\\
 & OGR & 333 & 0  & 42 & 125 & \textcolor[rgb]{0.08,0.5,0.24}{+42}\\
\hline
\multirow{3}{*}{Moatless}
 & UR  & 326 & 28 & 24 & 119 & \textcolor[rgb]{0.72,0.11,0.11}{-4} \\
 & JGR & 330 & 24 & 18 & 125 & \textcolor[rgb]{0.72,0.11,0.11}{-6} \\
 & OGR & 354 & 0  & 24 & 119 & \textcolor[rgb]{0.08,0.5,0.24}{+24} \\
\hline
\multirow{3}{*}{Openhands}
 & UR  & 340 & 19 & 23 & 116 & \textcolor[rgb]{0.08,0.5,0.24}{+4} \\
 & JGR & 343 & 16 & 22 & 117 & \textcolor[rgb]{0.08,0.5,0.24}{+6} \\
 & OGR & 359 & 0  & 23 & 116 & \textcolor[rgb]{0.08,0.5,0.24}{+23} \\
\hline
\end{tabular}
\end{table}

\textbf{\textit{Refinement Gains and Losses.}}
Table~\ref{disc_transition} decomposes the RQ-2 resolution delta into its four transitions.
Most importantly, refinement is not purely subtractive, since across all hosts and modes the refiner consistently recovers previously failing instances (I$\rightarrow$C ranges from 18 to 43), meaning refining a verbose patch can rewrite edits that actively prevented it from passing the test suite.
Recovery is strongest on SWE-agent (42–43 instances), whose extreme verbosity (+690.91\% total changes) makes extraneous edits most likely to interfere with correctness, demonstrating that verbosity is a correctness risk, not merely a review burden.


\begin{figure}[h]
\centerline{\includegraphics[width=.80\linewidth]{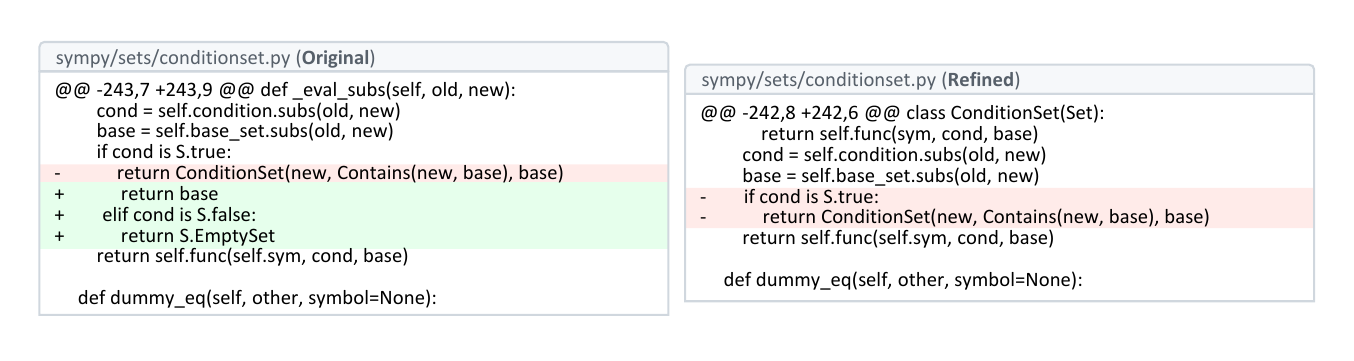}}
\caption{An example of patch refinement by only removal.}
\label{fig:ex_rm}
\end{figure}

\begin{figure}[t]
\centerline{\includegraphics[width=.80\linewidth]{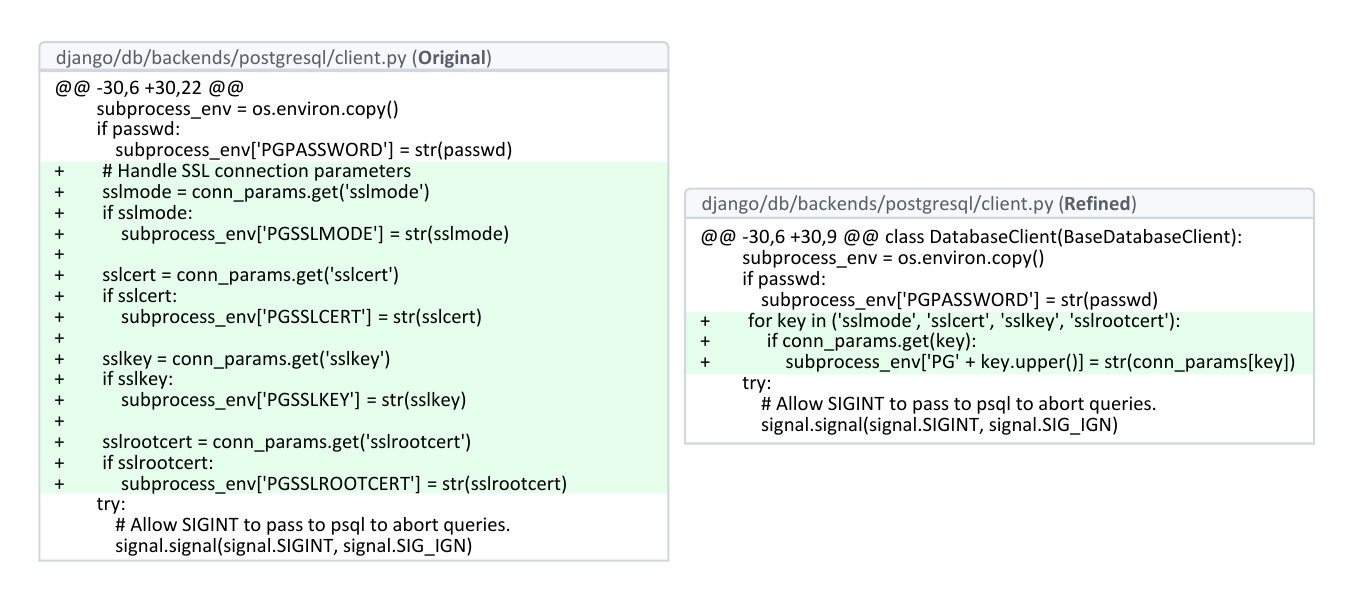}}
\caption{An example of fine-grained patch refinement.}
\label{fig:ex_fg}
\end{figure}

\textbf{\textit{Fine-grained Refinement Beyond Edit Removal.}}
A natural concern is that the refiner merely deletes lines, which would explain LOC reduction but not the cyclomatic-complexity drops in RQ-2.
Figures~\ref{fig:ex_rm} and~\ref{fig:ex_fg} show RECAP refining at two distinct levels.
In the removal case (Fig.~\ref{fig:ex_rm}), the candidate introduced an extra \texttt{if cond is S.true} scaffolding branch that the refiner recognized as redundant and dropped, keeping only the essential edit.
The fine-grained case (Fig.~\ref{fig:ex_fg}) is harder, where the candidate handled four SSL parameters with four near-identical \texttt{if} blocks and the refiner preserved exact behavior while collapsing them into a single loop.

%% file: 7.threats_to_validity.tex
\section{Threats to Validity}
\label{sec:threats}

\textbf{Internal validity.}
RQ-1 relies on public leaderboard artifacts and system descriptions, so undocumented settings may affect our classification of design choices.
We fix the collection date, use the latest submission per approach, classify from available papers, reports, and repositories, and treat the design-factor analysis as \emph{association} rather than \emph{causation}.
The pipeline's LLM-based judging, inflation, rejection sampling, and trace collection may introduce noisy supervision, but mitigated by combining multiple data sources and evaluating on real APR patches rather than synthetic inputs alone.

\textbf{Construct validity.}
SWE-bench tests serve as the executable oracle, so passing them shows plausibility under the benchmark rather than full semantic correctness.
We scope correctness claims to the evaluated resolution outcome and anchor OGR to the test suite, accepting a refined patch only when it preserves the host's passing outcome.
Our patch-size and complexity metrics are proxies for review effort, so we report three complementary perspectives spanning textual size, edit scope, and code complexity, all grounded in prior repair and review studies, and compare each patch against its developer patch on the same instance rather than absolute sizes.
Since developer patches are not guaranteed  to be globally minimal, we use them only as relative reference points.

\textbf{External validity.}
Our evaluation on SWE-bench may not transfer to other benchmarks, languages, proprietary repositories, or settings without tests.
Several factors limit this risk.
SWE-bench Verified spans real repository-level issues across many projects, and we evaluate across four host systems covering workflow-based and agent-based designs and three LLMs of different architectures and scales, so the gains are not tied to a single host or backbone.
The collector's framework-adaptive extraction also lowers the cost of porting the adapter elsewhere.
We scope conclusions to repository-level APR settings similar to SWE-bench, where candidate patches and validation outcomes are observable, and leave broader languages and settings to future work.

%% file: 8.conclusion.tex
\section{Conclusion}
\label{sec:conclusion}

This paper studied patch verbosity as a major concern in LLM-based APR.
Characterizing 28 SOTA approaches on SWE-bench Verified, we found even successful patches are consistently larger, broader, and structurally more complex, with verbosity tied to capability-oriented design choices rather than surface controls.
We therefore formulated post-generation patch refinement and proposed RECAP, a lightweight, plug-and-play adapter that attaches to existing frameworks after generation, trained on a multi-source dataset with SFT and DPO.
Across four host systems, it achieved a better size-correctness tradeoff than prompting, commit-untangling, and minimality-aware baselines, reducing edit volume, scope, and complexity toward or below gold levels while preserving or improving resolution.

%% file: acknowledgement.tex

%% file: main.bbl
\begin{thebibliography}{10}
\providecommand{\url}[1]{#1}
\csname url@samestyle\endcsname
\providecommand{\newblock}{\relax}
\providecommand{\bibinfo}[2]{#2}
\providecommand{\BIBentrySTDinterwordspacing}{\spaceskip=0pt\relax}
\providecommand{\BIBentryALTinterwordstretchfactor}{4}
\providecommand{\BIBentryALTinterwordspacing}{\spaceskip=\fontdimen2\font plus
\BIBentryALTinterwordstretchfactor\fontdimen3\font minus \fontdimen4\font\relax}
\providecommand{\BIBforeignlanguage}[2]{{%
\expandafter\ifx\csname l@#1\endcsname\relax
\typeout{** WARNING: IEEEtran.bst: No hyphenation pattern has been}%
\typeout{** loaded for the language `#1'. Using the pattern for}%
\typeout{** the default language instead.}%
\else
\language=\csname l@#1\endcsname
\fi
#2}}
\providecommand{\BIBdecl}{\relax}
\BIBdecl

\bibitem{pereira2025systematic}
A.~F. Pereira and R.~F. Mello, ``A systematic literature review on large language models applications in computer programming teaching evaluation process,'' \emph{IEEE Access}, 2025.

\bibitem{yang2025patch}
Y.~Yang, C.~Li, Z.~Han, R.~Li, K.~Xu, Q.~Li, W.~Zhong, Z.~Shen, Z.~Fei, J.~Ge \emph{et~al.}, ``Patch generation in apr: A survey from the perspectives of utilizing llms and using apr-specific information,'' \emph{ACM Transactions on Software Engineering and Methodology}, 2025.

\bibitem{yang2025survey}
B.~Yang, Z.~Cai, F.~Liu, B.~Le, L.~Zhang, T.~F. Bissyand{\'e}, Y.~Liu, and H.~Tian, ``A survey of llm-based automated program repair: Taxonomies, design paradigms, and applications,'' \emph{arXiv preprint arXiv:2506.23749}, 2025.

\bibitem{chen2024large}
Y.~Chen, J.~Wu, X.~Ling, C.~Li, Z.~Rui, T.~Luo, and Y.~Wu, ``When large language models confront repository-level automatic program repair: How well they done?'' in \emph{Proceedings of the 2024 IEEE/ACM 46th international conference on software engineering: companion proceedings}, 2024, pp. 459--471.

\bibitem{jiang2025agentic}
Z.~Jiang, D.~Lo, and Z.~Liu, ``Agentic software issue resolution with large language models: A survey,'' \emph{arXiv preprint arXiv:2512.22256}, 2025.

\bibitem{jimenez2024swe}
C.~E. Jimenez, J.~Yang, A.~Wettig, S.~Yao, K.~Pei, O.~Press, and K.~Narasimhan, ``Swe-bench: Can language models resolve real-world github issues?'' in \emph{International Conference on Learning Representations}, vol. 2024, 2024, pp. 54\,107--54\,157.

\bibitem{le2019automated}
C.~Le~Goues, M.~Pradel, and A.~Roychoudhury, ``Automated program repair,'' \emph{Communications of the ACM}, vol.~62, no.~12, pp. 56--65, 2019.

\bibitem{weissgerber2008small}
P.~Wei{\ss}gerber, D.~Neu, and S.~Diehl, ``Small patches get in!'' in \emph{Proceedings of the 2008 international working conference on Mining software repositories}, 2008, pp. 67--76.

\bibitem{rondon2025evaluating}
P.~Rondon, R.~Wei, J.~Cambronero, J.~Cito, A.~Sun, S.~Sanyam, M.~Tufano, and S.~Chandra, ``Evaluating agent-based program repair at google,'' in \emph{2025 IEEE/ACM 47th International Conference on Software Engineering: Software Engineering in Practice (ICSE-SEIP)}.\hskip 1em plus 0.5em minus 0.4em\relax IEEE, 2025, pp. 365--376.

\bibitem{tao2014writing}
Y.~Tao, D.~Han, and S.~Kim, ``Writing acceptable patches: An empirical study of open source project patches,'' in \emph{2014 IEEE International Conference on Software Maintenance and Evolution}.\hskip 1em plus 0.5em minus 0.4em\relax IEEE, 2014, pp. 271--280.

\bibitem{ebert2021exploratory}
F.~Ebert, F.~Castor, N.~Novielli, and A.~Serebrenik, ``An exploratory study on confusion in code reviews,'' \emph{Empirical Software Engineering}, vol.~26, no.~1, p.~12, 2021.

\bibitem{purushothaman2005toward}
R.~Purushothaman and D.~E. Perry, ``Toward understanding the rhetoric of small source code changes,'' \emph{IEEE Transactions on Software Engineering}, vol.~31, no.~6, pp. 511--526, 2005.

\bibitem{mechtaev2015directfix}
S.~Mechtaev, J.~Yi, and A.~Roychoudhury, ``Directfix: Looking for simple program repairs,'' in \emph{2015 IEEE/ACM 37th IEEE International Conference on Software Engineering}, vol.~1.\hskip 1em plus 0.5em minus 0.4em\relax IEEE, 2015, pp. 448--458.

\bibitem{li2025aligning}
J.~Li, H.~Zhu, H.~Liu, X.~Shi, H.~Zong, Y.~Dong, K.~Zhang, S.~Jiang, Z.~Jin, and G.~Li, ``Aligning llms to fully utilize the cross-file context in repository-level code completion,'' in \emph{2025 40th IEEE/ACM International Conference on Automated Software Engineering (ASE)}.\hskip 1em plus 0.5em minus 0.4em\relax IEEE, 2025, pp. 1477--1489.

\bibitem{tian2020evaluating}
H.~Tian, K.~Liu, A.~K. Kabor{\'e}, A.~Koyuncu, L.~Li, J.~Klein, and T.~F. Bissyand{\'e}, ``Evaluating representation learning of code changes for predicting patch correctness in program repair,'' in \emph{Proceedings of the 35th IEEE/ACM international conference on automated software engineering}, 2020, pp. 981--992.

\bibitem{li2022utango}
Y.~Li, S.~Wang, and T.~N. Nguyen, ``Utango: untangling commits with context-aware, graph-based, code change clustering learning model,'' in \emph{Proceedings of the 30th ACM Joint European Software Engineering Conference and Symposium on the Foundations of Software Engineering}, 2022, pp. 221--232.

\bibitem{zhu2026atomizer}
K.~Zhu, Z.~Tian, S.~Wang, M.~Leng, and X.~Mao, ``Atomizer: An llm-based collaborative multi-agent framework for intent-driven commit untangling,'' \emph{arXiv preprint arXiv:2601.01233}, 2026.

\bibitem{zhang2025toward}
M.~Zhang, Z.~Xu, Y.~Tian, X.~Cheng, and C.~Sun, ``Toward a better understanding of probabilistic delta debugging,'' in \emph{2025 IEEE/ACM 47th International Conference on Software Engineering (ICSE)}.\hskip 1em plus 0.5em minus 0.4em\relax IEEE, 2025, pp. 2024--2035.

\bibitem{zhou2025wdd}
X.~Zhou, Z.~Xu, M.~Zhang, Y.~Tian, and C.~Sun, ``Wdd: Weighted delta debugging,'' in \emph{2025 IEEE/ACM 47th International Conference on Software Engineering (ICSE)}.\hskip 1em plus 0.5em minus 0.4em\relax IEEE, 2025, pp. 1592--1603.

\bibitem{dai2025less}
Z.~Dai, B.~Chen, Z.~Zhao, X.~Tang, S.~Wu, C.~Yao, Z.~Gao, and J.~Chen, ``Less is more: Adaptive program repair with bug localization and preference learning,'' in \emph{Proceedings of the AAAI Conference on Artificial Intelligence}, vol.~39, no.~1, 2025, pp. 128--136.

\bibitem{ke2026qimeng}
C.~Ke, R.~Zhang, J.~Guo, Y.~Wen, L.~Ding, S.~Wang, X.~Zhu, X.~Peng, D.~Huang, Z.~Du \emph{et~al.}, ``Qimeng-prepair: Precise code repair via edit-aware reward optimization,'' \emph{arXiv preprint arXiv:2604.05963}, 2026.

\bibitem{swebenchlead}
SWE-bench, ``Swe-bench leaderboards,'' \url{https://www.swebench.com/}, 2025.

\bibitem{rafailov2023direct}
R.~Rafailov, A.~Sharma, E.~Mitchell, C.~D. Manning, S.~Ermon, and C.~Finn, ``Direct preference optimization: Your language model is secretly a reward model,'' \emph{Advances in neural information processing systems}, vol.~36, pp. 53\,728--53\,741, 2023.

\bibitem{feng2026empirical}
Y.~Feng and C.~Yang, ``An empirical study of sft-dpo interaction and parameterization in small language models,'' \emph{arXiv preprint arXiv:2603.20100}, 2026.

\bibitem{wei2022chain}
J.~Wei, X.~Wang, D.~Schuurmans, M.~Bosma, F.~Xia, E.~Chi, Q.~V. Le, D.~Zhou \emph{et~al.}, ``Chain-of-thought prompting elicits reasoning in large language models,'' \emph{Advances in neural information processing systems}, vol.~35, pp. 24\,824--24\,837, 2022.

\bibitem{zhuang2025unicott}
X.~Zhuang, Z.~Zhu, Z.~Wang, X.~Cheng, and Y.~Zou, ``Unicott: A unified framework for structural chain-of-thought distillation,'' in \emph{The Thirteenth International Conference on Learning Representations}, 2025.

\bibitem{yu2025cot}
P.~Yu, J.~Lanchantin, T.~Wang, W.~Yuan, O.~Golovneva, I.~Kulikov, S.~Sukhbaatar, J.~Weston, and J.~Xu, ``Cot-self-instruct: Building high-quality synthetic prompts for reasoning and non-reasoning tasks,'' \emph{arXiv preprint arXiv:2507.23751}, 2025.

\bibitem{sweagent}
J.~Yang, C.~E. Jimenez, A.~Wettig, K.~Lieret, S.~Yao, K.~Narasimhan, and O.~Press, ``Swe-agent: Agent-computer interfaces enable automated software engineering,'' \emph{Advances in Neural Information Processing Systems}, vol.~37, pp. 50\,528--50\,652, 2024.

\bibitem{wang2025openhands}
X.~Wang, S.~Rosenberg, J.~Michelini, C.~Smith, H.~Tran, E.~Nyst, R.~Malhotra, X.~Zhou, V.~Chen, R.~Brennan \emph{et~al.}, ``The openhands software agent sdk: A composable and extensible foundation for production agents,'' \emph{arXiv preprint arXiv:2511.03690}, 2025.

\bibitem{xia2024agentless}
C.~S. Xia, Y.~Deng, S.~Dunn, and L.~Zhang, ``Agentless: Demystifying llm-based software engineering agents,'' \emph{arXiv preprint arXiv:2407.01489}, 2024.

\bibitem{zhang2024autocoderover}
Y.~Zhang, H.~Ruan, Z.~Fan, and A.~Roychoudhury, ``Autocoderover: Autonomous program improvement,'' in \emph{Proceedings of the 33rd ACM SIGSOFT International Symposium on Software Testing and Analysis}, 2024, pp. 1592--1604.

\bibitem{moatless}
A.~Antoniades, A.~{\"O}rwall, K.~Zhang, Y.~Xie, A.~Goyal, and W.~Wang, ``Swe-search: Enhancing software agents with monte carlo tree search and iterative refinement,'' in \emph{International Conference on Learning Representations}, vol. 2025, 2025, pp. 64\,485--64\,515.

\bibitem{lingma}
Y.~Ma, Q.~Yang, R.~Cao, B.~Li, F.~Huang, and Y.~Li, ``Alibaba lingmaagent: Improving automated issue resolution via comprehensive repository exploration,'' in \emph{Proceedings of the 33rd ACM International Conference on the Foundations of Software Engineering}, 2025, pp. 238--249.

\bibitem{swefixer}
C.~Xie, B.~Li, C.~Gao, H.~Du, W.~Lam, D.~Zou, and K.~Chen, ``Swe-fixer: Training open-source llms for effective and efficient github issue resolution, 2025,'' \emph{URL https://arxiv. org/abs/2501.05040}, vol.~1, no.~2, p.~3.

\bibitem{deepswe2025}
M.~Luo, N.~Jain, J.~Singh, S.~Tan, A.~Patel, Q.~Wu, A.~Ariyak, C.~Cai, T.~Venkat, S.~Zhu, B.~Athiwaratkun, M.~Roongta, C.~Zhang, L.~E. Li, R.~A. Popa, K.~Sen, and I.~Stoica, ``Deepswe: Training a state-of-the-art coding agent from scratch by scaling rl,'' 2025, notion Blog.

\bibitem{sweexp}
S.~Chen, S.~Lin, Y.~Shi, H.~Lian, X.~Gu, L.~Yun, D.~Chen, L.~Cao, J.~Liu, N.~Xia \emph{et~al.}, ``Swe-exp: Experience-driven software issue resolution,'' \emph{arXiv preprint arXiv:2507.23361}, 2025.

\bibitem{yang2025lingxi}
X.~Yang, J.~Zhou, M.~Pacheco, W.~Zhu, P.~He, S.~Wang, K.~Liu, and R.~Pan, ``Lingxi: Repository-level issue resolution framework enhanced by procedural knowledge guided scaling,'' \emph{arXiv preprint arXiv:2510.11838}, 2025.

\bibitem{zeng2025skywork}
L.~Zeng, Y.~Li, Y.~Xiao, C.~Li, C.~Y. Liu, R.~Yan, T.~Wei, J.~He, X.~Song, Y.~Liu \emph{et~al.}, ``Skywork-swe: Unveiling data scaling laws for software engineering in llms,'' \emph{arXiv preprint arXiv:2506.19290}, 2025.

\bibitem{li2025patchpilot}
H.~Li, Y.~Tang, S.~Wang, and W.~Guo, ``Patchpilot: A cost-efficient software engineering agent with early attempts on formal verification,'' \emph{arXiv preprint arXiv:2502.02747}, 2025.

\bibitem{partachi2020flexeme}
\BIBentryALTinterwordspacing
P.-P. P{\^a}rțachi, S.~K. Dash, M.~Allamanis, and E.~T. Barr, ``Flexeme: Untangling commits using lexical flows,'' in \emph{Proceedings of the 28th ACM Joint European Software Engineering Conference and Symposium on the Foundations of Software Engineering (ESEC/FSE '20)}.\hskip 1em plus 0.5em minus 0.4em\relax ACM, 2020, pp. 63--74. [Online]. Available: \url{https://doi.org/10.1145/3368089.3409693}
\BIBentrySTDinterwordspacing

\bibitem{perses}
C.~Sun, Y.~Li, Q.~Zhang, T.~Gu, and Z.~Su, ``Perses: Syntax-guided program reduction,'' in \emph{Proceedings of the 40th International Conference on Software Engineering}.\hskip 1em plus 0.5em minus 0.4em\relax Association for Computing Machinery, 2018, p. 361–371.

\bibitem{SongWLc2d2}
\BIBentryALTinterwordspacing
X.~Song, Y.~Wu, S.~Liu, B.~C. 0001, Y.~L. 0001, and X.~P. 0001, ``C2d2: Extracting critical changes for real-world bugs with dependency-sensitive delta debugging,'' in \emph{Proceedings of the 33rd ACM SIGSOFT International Symposium on Software Testing and Analysis, ISSTA 2024, Vienna, Austria, September 16-20, 2024}, M.~Christakis and M.~Pradel, Eds.\hskip 1em plus 0.5em minus 0.4em\relax ACM, 2024, pp. 300--312. [Online]. Available: \url{https://doi.org/10.1145/3650212.3652129}
\BIBentrySTDinterwordspacing

\bibitem{koh2026detecting}
B.~Koh, N.~Walkinshaw, and D.~Shin, ``Detecting multiple semantic concerns in tangled code commits,'' \emph{arXiv preprint arXiv:2601.21298}, 2026.

\bibitem{zeng2025first}
Q.~Zeng, Y.~Zhang, Z.~Qiu, and H.~Liu, ``A first look at conventional commits classification,'' in \emph{2025 IEEE/ACM 47th International Conference on Software Engineering (ICSE)}.\hskip 1em plus 0.5em minus 0.4em\relax IEEE, 2025, pp. 2277--2289.

\bibitem{cortexa}
A.~Sohrabizadeh, J.~Song, M.~Liu, R.~Roy, C.~Lee, J.~Raiman, and B.~Catanzaro, ``Nemotron-cortexa: Enhancing llm agents for software engineering tasks via improved localization and solution diversity,'' in \emph{Forty-second International Conference on Machine Learning}, 2025.

\bibitem{li2024llms}
H.~Li, Q.~Dong, J.~Chen, H.~Su, Y.~Zhou, Q.~Ai, Z.~Ye, and Y.~Liu, ``Llms-as-judges: a comprehensive survey on llm-based evaluation methods,'' \emph{arXiv preprint arXiv:2412.05579}, 2024.

\bibitem{li2025hybrid}
F.~Li, J.~Jiang, J.~Sun, and H.~Zhang, ``Hybrid automated program repair by combining large language models and program analysis,'' \emph{ACM Transactions on Software Engineering and Methodology}, vol.~34, no.~7, pp. 1--28, 2025.

\bibitem{liu2025empirical}
S.~Liu, F.~Liu, L.~Li, X.~Tan, Y.~Zhu, X.~Lian, and L.~Zhang, ``An empirical study on failures in automated issue solving,'' \emph{arXiv preprint arXiv:2509.13941}, 2025.

\bibitem{alomar2025chatgpt}
E.~A. AlOmar, L.~Xu, S.~Martinez, A.~Peruma, M.~W. Mkaouer, C.~D. Newman, and A.~Ouni, ``Chatgpt for code refactoring: Analyzing topics, interaction, and effective prompts,'' in \emph{2025 IEEE International Conference on Collaborative Advances in Software and COmputiNg (CASCON)}.\hskip 1em plus 0.5em minus 0.4em\relax IEEE, 2025, pp. 389--398.

\bibitem{xu2025aligning}
J.~Xu, Y.~Fu, S.~H. Tan, and P.~He, ``Aligning the objective of llm-based program repair,'' in \emph{2025 IEEE/ACM 47th International Conference on Software Engineering (ICSE)}.\hskip 1em plus 0.5em minus 0.4em\relax IEEE, 2025, pp. 2548--2560.

\bibitem{campos2025empirical}
V.~Campos, R.~Shariffdeen, A.~Ulges, and Y.~Noller, ``Empirical evaluation of generalizable automated program repair with large language models,'' \emph{arXiv preprint arXiv:2506.03283}, 2025.

\bibitem{zawalski2024robotic}
M.~Zawalski, W.~Chen, K.~Pertsch, O.~Mees, C.~Finn, and S.~Levine, ``Robotic control via embodied chain-of-thought reasoning,'' \emph{arXiv preprint arXiv:2407.08693}, 2024.

\bibitem{li2025llms}
D.~Li, S.~Cao, T.~Griggs, S.~Liu, X.~Mo, E.~Tang, S.~Hegde, K.~Hakhamaneshi, S.~G. Patil, M.~Zaharia \emph{et~al.}, ``Llms can easily learn to reason from demonstrations structure, not content, is what matters!'' \emph{arXiv preprint arXiv:2502.07374}, 2025.

\bibitem{xu2026mind}
H.~Xu, Y.~Yan, Y.~Shen, W.~Zhang, G.~Hou, S.~Jiang, K.~Song, W.~Lu, J.~Xiao, and Y.~Zhuang, ``Mind the gap: Bridging thought leap for improved chain-of-thought tuning,'' \emph{Advances in Neural Information Processing Systems}, vol.~38, pp. 143\,665--143\,694, 2026.

\bibitem{yang2025semantics}
C.~Yang, T.~Zhang, J.~Jiang, X.~Zhou, H.~Tian, M.~Du, J.~Shi, J.~Chen, Y.~Li, E.~L. Ouh \emph{et~al.}, ``Semantics-aligned, curriculum-driven, and reasoning-enhanced vulnerability repair framework,'' \emph{arXiv preprint arXiv:2510.01002}, 2025.

\bibitem{yin2026improving}
S.~Yin, Z.~Tian, J.~Chen, and S.~Guo, ``Improving llm code generation via requirement-aware curriculum reinforcement learning,'' \emph{arXiv preprint arXiv:2605.00433}, 2026.

\bibitem{schulman2017proximal}
J.~Schulman, F.~Wolski, P.~Dhariwal, A.~Radford, and O.~Klimov, ``Proximal policy optimization algorithms,'' \emph{arXiv preprint arXiv:1707.06347}, 2017.

\bibitem{shao2024deepseekmath}
Z.~Shao, P.~Wang, Q.~Zhu, R.~Xu, J.~Song, X.~Bi, H.~Zhang, M.~Zhang, Y.~Li, Y.~Wu \emph{et~al.}, ``Deepseekmath: Pushing the limits of mathematical reasoning in open language models,'' \emph{arXiv preprint arXiv:2402.03300}, 2024.

\bibitem{gpt52}
OpenAI, ``Introducing gpt-5.2 | openai,'' \url{https://openai.com/index/introducing-gpt-5-2/}, 2025.

\bibitem{gemini3}
Google, ``Introducing our most intelligent model yet. with state-of-the-art reasoning to help you learn, build, and plan anything.'' \url{https://deepmind.google/models/gemini/}, 2025.

\bibitem{qwen3.5}
\BIBentryALTinterwordspacing
{Qwen Team}, ``{Qwen3.5}: Towards native multimodal agents,'' February 2026. [Online]. Available: \url{https://qwen.ai/blog?id=qwen3.5}
\BIBentrySTDinterwordspacing

\bibitem{cs45}
Anthropic, ``Introducing claude sonnet 4.5,'' \url{https://www.anthropic.com/news/claude-sonnet-4-5}, 2025.

\bibitem{gemma4}
G.~Deepmind, ``Our most intelligent open models, built from gemini 3 research and technology to maximize intelligence-per-parameter,'' \url{https://deepmind.google/models/gemma/gemma-4/}, 2026.

\bibitem{mistral3}
M.~AI, ``Introducing mistral 3 | mistral ai,'' \url{https://mistral.ai/news/mistral-3}, 2025.

\bibitem{zhang2015interactive}
T.~Zhang, M.~Song, J.~Pinedo, and M.~Kim, ``Interactive code review for systematic changes,'' in \emph{2015 IEEE/ACM 37th IEEE International Conference on Software Engineering}, vol.~1.\hskip 1em plus 0.5em minus 0.4em\relax IEEE, 2015, pp. 111--122.

\bibitem{shariffdeen2020automated}
R.~S. Shariffdeen, S.~H. Tan, M.~Gao, and A.~Roychoudhury, ``Automated patch transplantation,'' \emph{ACM Transactions on Software Engineering and Methodology (TOSEM)}, vol.~30, no.~1, pp. 1--36, 2020.

\bibitem{thongtanunam2017review}
P.~Thongtanunam, S.~McIntosh, A.~E. Hassan, and H.~Iida, ``Review participation in modern code review: An empirical study of the android, qt, and openstack projects,'' \emph{Empirical Software Engineering}, vol.~22, no.~2, pp. 768--817, 2017.

\bibitem{baysal2016investigating}
O.~Baysal, O.~Kononenko, R.~Holmes, and M.~W. Godfrey, ``Investigating technical and non-technical factors influencing modern code review,'' \emph{Empirical Software Engineering}, vol.~21, no.~3, pp. 932--959, 2016.

\bibitem{gitdiff}
Git, ``Git - git-diff documentation,'' \url{https://git-scm.com/docs/git-diff}, 2026.

\bibitem{gao2025trae}
P.~Gao, Z.~Tian, X.~Meng, X.~Wang, R.~Hu, Y.~Xiao, Y.~Liu, Z.~Zhang, J.~Chen, C.~Gao \emph{et~al.}, ``Trae agent: An llm-based agent for software engineering with test-time scaling,'' \emph{arXiv preprint arXiv:2507.23370}, 2025.

\bibitem{swerizzo}
SWE-Rizzo, ``Swe-rizzo - software engineering on bittensor - a team rizzo subnet,'' \url{https://github.com/brokespace/code}, 2025.

\bibitem{CodeSweep}
CodeSweep, ``Codesweep - autopilot for enterprise software maintenance,'' \url{https://codesweep.ai/}, 2025.

\bibitem{JoyCode}
JD, ``Joycode swe-bench agent pipeline,'' \url{https://github.com/jd-opensource/joycode-agent/}, 2025.

\bibitem{Refact}
S.~M.~C. Ai, ``Refact,'' \url{https://github.com/smallcloudai/refact}, 2025.

\bibitem{composio}
Composio, ``Composio sdk,'' \url{https://github.com/ComposioHQ/composio}, 2024.

\bibitem{navie}
AppMap, ``Appmap navie,'' \url{https://appmap.io/product/appmap-navie.html}, 2024.

\bibitem{augment}
AugmentCode, ``Augment swe-bench verified agent,'' \url{https://github.com/augmentcode/augment-swebench-agent}, 2025.

\bibitem{Zai}
Z.ai, ``Z.ai: Free ai chatbot,'' \url{https://chat.z.ai/}, 2025.

\bibitem{gru}
Gru, ``Gru ai - your agi assistant,'' \url{https://gru.ai/}, 2024.

\bibitem{entropo}
J.~Yu, Z.~Cheng, X.~Wu, and X.~Xing, ``Building coding agents via entropy-enhanced multi-turn preference optimization,'' \emph{arXiv preprint arXiv:2509.12434}, 2025.

\bibitem{halstead1977elements}
M.~H. Halstead, \emph{Elements of Software Science (Operating and programming systems series)}.\hskip 1em plus 0.5em minus 0.4em\relax Elsevier Science Inc., 1977.

\bibitem{ebert2016cyclomatic}
C.~Ebert, J.~Cain, G.~Antoniol, S.~Counsell, and P.~Laplante, ``Cyclomatic complexity,'' \emph{IEEE software}, vol.~33, no.~6, pp. 27--29, 2016.

\bibitem{gulwani2018automated}
S.~Gulwani, I.~Radi{\v{c}}ek, and F.~Zuleger, ``Automated clustering and program repair for introductory programming assignments,'' \emph{ACM SIGPLAN Notices}, vol.~53, no.~4, pp. 465--480, 2018.

\bibitem{mann1947test}
H.~B. Mann and D.~R. Whitney, ``On a test of whether one of two random variables is stochastically larger than the other,'' \emph{The annals of mathematical statistics}, pp. 50--60, 1947.

\bibitem{wilcoxon1992individual}
F.~Wilcoxon, ``Individual comparisons by ranking methods,'' in \emph{Breakthroughs in statistics: Methodology and distribution}.\hskip 1em plus 0.5em minus 0.4em\relax Springer, 1992, pp. 196--202.

\bibitem{ostertagova2014methodology}
E.~Ostertagova, O.~Ostertag, and J.~Kov{\'a}{\v{c}}, ``Methodology and application of the kruskal-wallis test,'' \emph{Applied mechanics and materials}, vol. 611, pp. 115--120, 2014.

\bibitem{holm1979simple}
S.~Holm, ``A simple sequentially rejective multiple test procedure,'' \emph{Scandinavian journal of statistics}, pp. 65--70, 1979.

\bibitem{cliff1993dominance}
N.~Cliff, ``Dominance statistics: Ordinal analyses to answer ordinal questions.'' \emph{Psychological bulletin}, vol. 114, no.~3, p. 494, 1993.

\end{thebibliography}
